\documentclass[aps,prd,onecolumn,showpacs,groupedaddress]{revtex4-2}
\usepackage{graphicx}
\usepackage{dcolumn}
\usepackage{bm}
\usepackage{color}
\usepackage{longtable}
\usepackage{supertabular}
\usepackage[T1]{fontenc}
\usepackage{epstopdf}
\usepackage{amsmath}
\usepackage{amssymb}
\usepackage{setspace}
\usepackage{multirow}
\usepackage{booktabs}
\usepackage[section]{placeins}
\usepackage{mathrsfs}
\usepackage{hyperref}
\usepackage{relsize}

\makeatletter

\newcommand{\Rmnum}[1]{\expandafter\@slowromancap\romannumeral #1@} 
\newcommand{\bq}{\begin{equation}}
\newcommand{\eq}{\end{equation}}
\newcommand{\bqn}{\begin{eqnarray}}
\newcommand{\eqn}{\end{eqnarray}}
\newcommand{\nb}{\nonumber}

\makeatother

\begin{document}

\title{Effectiveness of nonflow suppression using multi-particle correlators}

\author{Chong Ye$^{2, 1}$}
\author{Wei-Liang Qian$^{3, 4, 1}$}\email[E-mail: ]{wlqian@usp.br (corresponding author)}
\author{Yue Cui$^{2}$}
\author{Dan Wen$^{5}$}
\author{Yutao Xing$^{2}$}\email[E-mail: ]{xy@id.uff.br (corresponding author)}
\author{Rui-Hong Yue$^{1}$}\email[E-mail: ]{rhyue@yzu.edu.cn (corresponding author)}
\author{Takeshi Kodama$^{6, 2}$}

\affiliation{$^{1}$ Center for Gravitation and Cosmology, School of Physical Science and Technology, Yangzhou University, 225009, Yangzhou, China}
\affiliation{$^{2}$ Instituto de F\'isica, Universidade Federal Fluminense, 24210-346, Niter\'oi, RJ, Brazil}
\affiliation{$^{3}$ Escola de Engenharia de Lorena, Universidade de S\~ao Paulo, 12602-810, Lorena, SP, Brazil}
\affiliation{$^{4}$ Faculdade de Engenharia de Guaratinguet\'a, Universidade Estadual Paulista, 12516-410, Guaratinguet\'a, SP, Brazil}
\affiliation{$^{5}$ Chongqing University of Posts and Telecommunications, Chongqing 400065, China}
\affiliation{$^{6}$ Instituto de F\'isica, Universidade Federal do Rio de Janeiro, 21945-970, Rio de Janeiro-RJ , Brazil}

\begin{abstract}
As flow estimators, multi-particle correlators, particularly the higher-order ones, are generally regarded as effective tools for suppressing non-flow contributions.
In this work, however, using two well-known toy models that simulate non-flow effects, we demonstrate that multi-particle correlators can, especially in small systems, yield estimates that deviate even further from the underlying flow harmonics than those obtained from other conventional approaches.
The two toy models considered here are designed to mimic non-flow effects arising from particle decay and global momentum conservation, such that the {\it apparent} harmonic coefficients become significantly different from the {\it input} values.
We provide an analytic explanation for the observed behavior of flow estimates based on multi-particle correlators.
Specifically, in the toy model mimicking particle decay, we elucidate the oscillations observed in $v_2\{2\}$ and $v_2\{4\}$.
For the other toy model simulating momentum conservation, we show that multi-particle cumulants introduce a deformation in the collective flow that is unique to multi-particle correlators.
Additionally, we compare these results with those obtained using the maximum-likelihood estimation method, a recently proposed flow estimator that serves as a viable alternative to traditional techniques.
\end{abstract}

\date{Dec. 27th, 2025}

\maketitle
\newpage

\section{Introduction}\label{section1}

The quark-gluon plasma (QGP)~\cite{qgp-review-12, qgp-review-13} is a deconfined state of strongly interacting matter in which quarks and gluons behave as nearly free color charges due to the asymptotic freedom of quantum chromodynamics.
Such an extremely hot and dense medium is transiently created in relativistic heavy-ion collisions at RHIC and the LHC~\cite{RHIC-star-overview-1, RHIC-brahms-overview-1, RHIC-phenix-overview-1, RHIC-phobos-overview-1, LHC-alice-review-01, LHC-atlas-review-01, LHC-cms-review-01}.
From a phenomenological perspective, relativistic hydrodynamics has emerged as the standard macroscopic framework to describe the space-time evolution of the QGP produced in these collisions~\cite{hydro-review-04, hydro-review-05, hydro-review-06, hydro-review-07, hydro-review-08, hydro-review-09, hydro-review-10}.
In this description, the medium is modeled as a nearly locally equilibrated continuum fluid whose collective expansion encodes the essential physics underlying the observed hadronic signals.
Key soft-sector observables sensitive to this collective behavior include identified-particle spectra at low and intermediate transverse momentum, anisotropic flow coefficients, and multi-particle correlations.
The measurement of strong azimuthal anisotropies at RHIC led to the characterization of the QGP as an almost {\it perfect} liquid with very small shear viscosity to entropy density ratio~\cite{RHIC-star-v2-01}.
Consequently, azimuthal flow coefficients have become indispensable probes for constraining transport properties and geometric features of the expanding system~\cite{RHIC-brahms-v2-01, RHIC-phenix-v2-01, RHIC-star-v2-05, LHC-alice-vn-01, LHC-atlas-vn-01, LHC-cms-vn-01}.
Investigations of collective flow in nuclear collisions have been carried out in particular for small collision systems~\cite{LHC-cms-vn-04, LHC-atlas-vn-07, LHC-small-system-review-02}, collisions involving deformed nuclei~\cite{RHIC-star-v2-10, hydro-vn-deformed-07, hydro-vn-deformed-10, hydro-vn-deformed-11, hydro-vn-deformed-12}, and, more recently, through detailed studies of radial flow fluctuations and their correlations~\cite{LHC-atlas-vn-20}.
Within the hydrodynamic paradigm, the collective evolution is primarily governed by the medium response to event-by-event fluctuations in the initial energy density profile.
The intrinsically nonlinear structure of the hydrodynamic equations has spurred extensive efforts to quantify how fluctuating initial conditions translate into final-state azimuthal anisotropies and their correlations~\cite{hydro-v3-02,hydro-vn-33,sph-vn-03,hydro-vn-34,sph-vn-04,sph-vn-06,hydro-vn-45, sph-v2-02, sph-corr-01, sph-corr-19, sph-corr-26, sph-corr-28, sph-corr-29, sph-corr-30,sph-corr-33}.

The collective flow is quantified in terms of the Fourier expansion of the one-particle distribution function
\begin{eqnarray}
f_1(\phi)=\frac{1}{2\pi}\left[1+\sum_{n=1}^{} 2v_{n}\cos{n(\phi-\Psi_n)}\right],
\label{oneParDis}
\end{eqnarray}
where $\phi$ denotes the azimuthal emission angle of the particle and $\Psi_n$ specifies the corresponding symmetry plane of order $n$.
The flow harmonics
\bqn
v_n = \langle \cos n(\phi-\Psi_n) \rangle = \int d\phi \cos n(\phi-\Psi_n) f_1(\phi) ,\label{defvn}
\eqn
are essentially the Fourier components that encode the degree of anisotropic collective motion~\cite{event-plane-method-1}.
In particular, the second-order coefficient $v_2$ is referred to as elliptic flow, and the third-order harmonic $v_3$ is known as triangular flow.

Regarding traditional approaches, the event-plane method infers $\Psi_n$ to evaluate the Fourier coefficients defined in Eq.~\eqref{oneParDis}~\cite{event-plane-method-1, event-plane-method-2}.
This construction effectively uses an experimental proxy for the reaction plane, which cannot be directly accessed in a heavy-ion collision.
An alternative strategy is provided by methods formulated in terms of particle correlations, typically expressed through flow Q-vectors and multi-particle cumulants~\cite{hydro-corr-ph-03, hydro-corr-ph-10, pythia-vn-10}.
A key advantage of such correlation techniques is that the explicit dependence on the event-plane angles in Eq.~\eqref{oneParDis} drops out, thereby yielding observables that do not require an explicit reconstruction of $\Psi_n$.
Furthermore, the cumulant hierarchy admits a compact representation in terms of a generating function, which makes the treatment of higher-order correlations more systematic.
Within this broad class of correlation-based approaches one finds the standard multi-particle cumulant method~\cite{hydro-corr-ph-03, hydro-corr-ph-04, hydro-corr-ph-10, hydro-corr-ph-23, hydro-corr-ph-27}, Lee-Yang-zero techniques~\cite{hydro-corr-LY-zeros-01, hydro-corr-LY-zeros-02}, and symmetric cumulants probing correlations between different flow harmonics~\cite{hydro-corr-ph-36}, along with several related refinements~\cite{hydro-corr-ph-38, hydro-corr-ph-42, hydro-vn-pca-01}.
It should be emphasized that in the notation of multi-particle cumulants, distinct definitions of flow coefficients naturally arise, depending on the order of the moment used to construct the corresponding estimator.
Recently, the maximum likelihood estimator (MLE) has been advocated as a conceptually different tool for extracting flow harmonics~\cite{sph-vn-10, sph-vn-11, sph-vn-12}.
In this formulation, the $v_n$ are treated as parameters of a postulated probability density for particle emission, and the MLE, a standard {\it estimator} in statistical inference~\cite{book-statistical-inference-Wasserman}, is employed to infer these parameters from the measured events.

An essential ingredient in this framework is the presence of non-flow contributions~\cite{hydro-corr-non-flow-01, hydro-corr-non-flow-04, hydro-corr-ph-09, hydro-corr-non-flow-review-03}, namely correlations that cannot be accounted for by independent particle emission from a single-particle distribution. 
Typical sources of non-flow include resonance decays~\cite{hydro-corr-non-flow-04}, jet fragmentation and showering~\cite{qgp-review-15, qgp-review-20}, string-breaking processes~\cite{jet-fragmentation-02}, Hanbury-Brown-Twiss quantum interference~\cite{hbt-20}, back-to-back jet-like correlations~\cite{bbc-02}, as well as a variety of final-state interactions. 
In addition, exact conservation laws such as global momentum conservation generate long-range azimuthal correlations that differ from those expected for a system governed purely by collective flow. 
When momentum conservation is implemented as a constraint on the multi-particle phase space, its impact on correlation observables can be approximated analytically, for instance by invoking the central limit theorem~\cite{hydro-corr-non-flow-01} or by employing saddle-point techniques~\cite{hydro-corr-non-flow-02}. 
Over the past decade, these ideas have motivated a range of studies aimed at quantifying and modeling non-flow effects in different kinematic regimes and collision systems~\cite{hydro-corr-non-flow-01, hydro-corr-non-flow-02, hydro-corr-non-flow-03, hydro-corr-non-flow-07, hydro-corr-non-flow-08, hydro-corr-non-flow-13, hydro-corr-non-flow-14, hydro-corr-non-flow-20}. 
For collisions of large nuclei at top RHIC and LHC energies, such contributions are often subleading, whereas explicit model calculations~\cite{hydro-corr-non-flow-14, hydro-corr-non-flow-20} indicate that non-flow can become sizable in small or low-multiplicity systems, where it may compete with or even dominate the collective signal. 
These observations have stimulated renewed interest in developing robust non-flow mitigation and estimation strategies in recent years~\cite{hydro-corr-non-flow-review-03}.

Notably, it is widely recognized that non-flow contributions are strongly reduced when higher-order multi-particle cumulants are employed~\cite{hydro-corr-ph-03, hydro-corr-ph-04}.
Heuristically, this suppression arises because the fraction of particle tuples that receive sizable contributions from non-flow sources, which are typically short range, decreases rapidly as the order of the correlator increases.
The influence of non-flow on flow measurements has been investigated explicitly through numerical simulations in a variety of model setups~\cite{hydro-corr-ph-03, hydro-corr-ph-04, RHIC-star-v2-07}.
In Refs.~\cite{hydro-corr-ph-03, hydro-corr-ph-04}, this was achieved by superimposing additional particle pairs with identical azimuthal angles on top of an underlying background exhibiting collective flow. 
A complementary study by the STAR Collaboration~\cite{RHIC-star-v2-07} considered correlated particle pairs separated by a fixed relative azimuthal angle and examined the resulting signatures.
In these analyses, the elliptic flow signal was evaluated as a function of the pair opening angle $\phi_\mathrm{open}$ using several commonly employed flow estimators.
The authors concluded that the four-particle estimate $v_2\{4\}$ provides a more robust determination of elliptic flow than more conventional two-particle-based measures in the presence of non-flow.
Global momentum conservation constitutes another important source of non-flow correlations. 
Its impact on azimuthal correlation observables has been systematically analyzed in a number of works~\cite{hydro-corr-non-flow-01, hydro-corr-non-flow-02, hydro-corr-non-flow-03, hydro-corr-non-flow-07, hydro-corr-non-flow-08, hydro-corr-non-flow-13, hydro-corr-non-flow-14, hydro-corr-non-flow-20}.
From an analytical standpoint, these studies compute corrections to the $k$-particle correlation functions under the assumption of independent particle emission from a common one-particle distribution, with transverse momentum conservation implemented as a global constraint. 
The expressions simplify considerably if one neglects background anisotropic flow and assumes an isotropic single-particle distribution.
At leading order in this limit, the correction is nonzero only for the directed flow coefficient $v_1\{2\}= \sqrt{c_1\{2\}}$, with ${c_1\{2\}}\propto 1/M$ for a system without collective flow~\cite{hydro-corr-non-flow-01, hydro-corr-non-flow-02}. 
This result has subsequently been generalized~\cite{hydro-corr-non-flow-13, hydro-corr-non-flow-20} to obtain the leading non-vanishing contribution $c_n\{2k\}\propto 1/(M-2k)^{nk}$ for higher-order multi-particle cumulants. 
When a realistic background flow is included, the induced corrections to the measured collective flow signals, though typically small, become more pronounced and exhibit a strong dependence on the specific correlator considered.
For elliptic flow, the leading corrections to the corresponding cumulants behave as $\Delta c_2\{2\}\propto 1/M$ and $\Delta c_2\{4\}\propto 1/M$~\cite{hydro-corr-non-flow-14}, and can be generalized to $\Delta c_n\{2k\}\propto 1/(M-2k)$~\cite{hydro-corr-non-flow-20}. 
In particular, it has been emphasized~\cite{hydro-corr-non-flow-14} that a competition between collective flow and non-flow contributions can drive a sign change of $c_2\{4\}$ as the event multiplicity is varied.
Nevertheless, the non-flow-induced corrections to both $c_2\{2\}$ and $c_2\{4\}$ remain positive in these scenarios.

The present work is motivated by these developments and aims to examine in detail how effectively non-flow contributions can be suppressed when particle correlators are used as flow estimators. 
To this end, a detailed numerical investigation is carried out using two illustrative toy models designed to mimic representative classes of non-flow effects. 
The performance of multi-particle correlators is then assessed, with comparisons made where appropriate to alternative approaches such as the event-plane method and the MLE estimator.
For certain scenarios, the particle-correlation estimators are found to deviate from the input flow harmonics, whereas other approaches yield results that remain comparatively close to the underlying background collective flow before it is distorted by non-flow.
Such differences are more pronounced in small or low-multiplicity systems, where non-flow contributions play a more prominent role and can effectively mask genuine collective behavior.

The remainder of this paper is organized as follows.
In the next section, we provide a brief review of two toy models that emulate non-flow effects associated with particle decay and global momentum conservation.
In Secs.~\ref{section3}, we present numerical results for flow extraction based on multi-particle correlators and compare their performance with that of alternative approaches.
In Sec.~\ref{section4}, we give an analytic account of the mechanisms responsible for the observed deviations from the input flow harmonics.
The final section summarizes the main findings and offers concluding remarks.

\section{Two toy models for non-flow effects}\label{section2}

To simulate the non-flow contributions, we employ two toy models that respectively mimic particle decays~\cite{hydro-corr-ph-03, hydro-corr-ph-04, RHIC-star-v2-07} and global momentum conservation~\cite{hydro-corr-non-flow-01, hydro-corr-non-flow-02, hydro-corr-non-flow-03, hydro-corr-non-flow-07, hydro-corr-non-flow-08, hydro-corr-non-flow-13, hydro-corr-non-flow-14, hydro-corr-non-flow-20}.
These toy models are constructed in a minimal fashion, with the resulting non-flow effects potentially being intentionally exaggerated.
For simplicity, the system is assumed to undergo Bjorken-invariant expansion, so that only the particles' azimuthal angles are taken into account.
Our aim is to examine how the non-flow suppression in the particle correlator compares with that obtained from other conventional approaches, and whether this necessarily leads to a better reconstruction of the underlying flow harmonics.

Toy model I accounts for particle decays.
An event first generates $N$ particles (with azimuthal angles denoted by $\phi$) emitted independently according to the one-particle distribution in Eq.~\eqref{oneParDis}, where the underlying flow harmonics are referred to as the {\it input} values.
The effect of the decay process is then modeled by the emission of $M$ additional particle pairs.
Following Refs.~\cite{hydro-corr-ph-03, hydro-corr-ph-04, RHIC-star-v2-07}, extra particle pairs with a prescribed opening angle $\phi_\mathrm{open}$ are embedded in the background particles.
Specifically, we consider two scenarios in which the emission of these particle pairs is either correlated or uncorrelated with the symmetry plane.
In the correlated case, the first particle is sampled from the distribution in Eq.~\eqref{oneParDis}, and the second particle is emitted at an azimuthal direction such that the pair forms a prescribed opening angle $\phi_\mathrm{open}$.
Subsequently, the resulting particle distribution implies flow harmonics that typically deviate from the original ones, which, as elaborated below, will be referred to as {\it apparent} values.

\begin{figure}[ht]
    \centering
    \begin{minipage}{0.32\textwidth}
        \centering
        \includegraphics[width=1.1\textwidth, height=0.26\textheight]{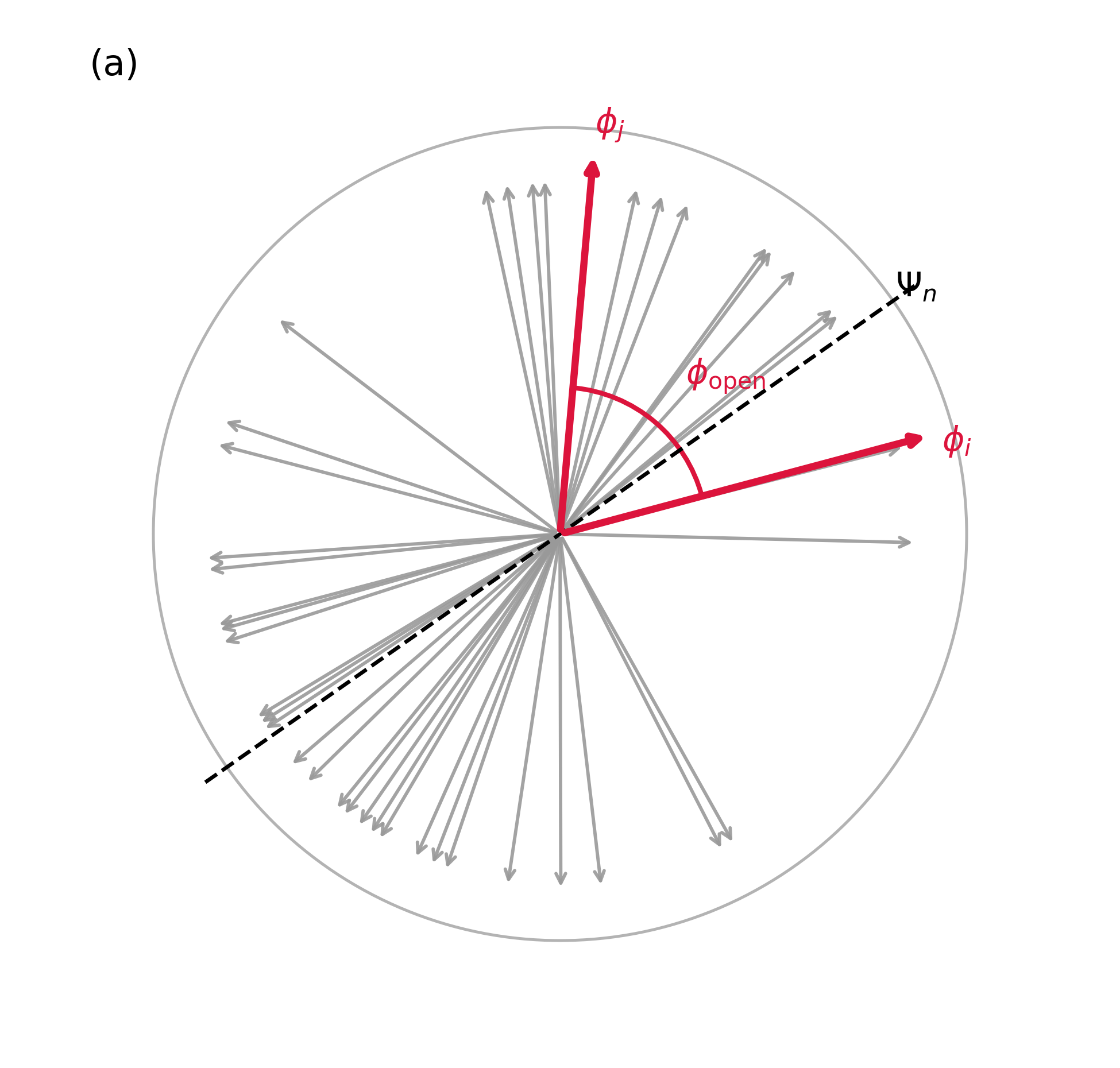}
    \end{minipage}
    \hspace{0.1\textwidth}
    \begin{minipage}{0.32\textwidth}
        \centering
        \includegraphics[width=1.1\textwidth, height=0.26\textheight]{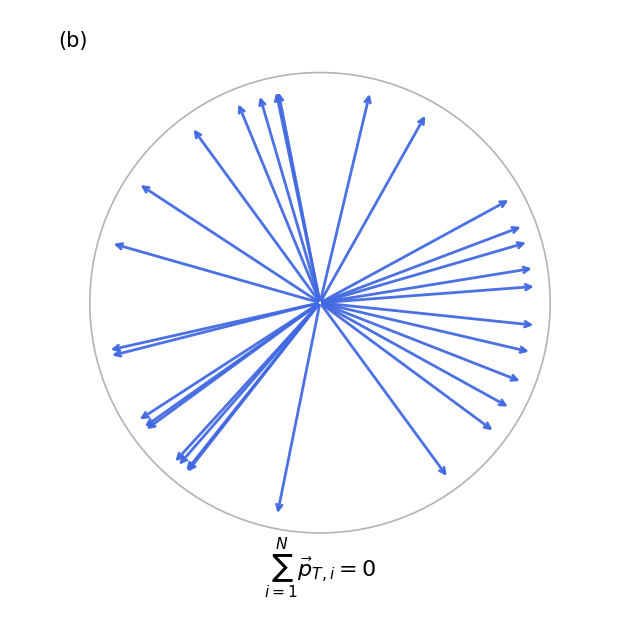}
    \end{minipage}
\renewcommand{\figurename}{Fig.}   
\caption{The two toy models employed in the present study.
Left: The first scenario represents particle decays.  
Particle pairs with an opening angle $\phi_\mathrm{open}$ are added on top of the background flow generated according to the one-particle distribution in Eq.~\eqref{oneParDis}.  
In the panel, $\phi_i$ and $\phi_j$ denote the azimuthal angles of the first and second particles forming a pair that models the non-flow contribution arising from particle decay, and the opening angle is defined as $\phi_\mathrm{open}=\phi_j-\phi_i$.  
Right: The second scenario incorporates global momentum conservation.
The particle emission follows the so-called T-generation algorithm~\cite{jet-ph-04}, which guarantees that global momentum conservation is imposed as a constraint on otherwise independent isotropic particle emission.}
\label{fig_toymodel_scheme}
\end{figure}

Toy model II mimics the effect of global momentum conservation.
Theoretically, global momentum conservation is imposed by constraining the multi-particle distribution function, originally governed by i.i.d.\ sampling from Eq.~\eqref{oneParDis}, to satisfy total momentum conservation.

Numerically, event generation under global momentum conservation is implemented as a Monte Carlo sampler that produces events according to a probability distribution satisfying these constraints.
An algorithm that implements global momentum conservation into an otherwise isotropic and independent particle emission is known as T-generation, which has been developed for a given isotropic single-particle distribution function.
This algorithm was implemented in the GENBOD code~\cite{jet-ph-04}.
In the present study, we employ an implementation provided by the \texttt{TGenPhaseSpace} class in ROOT~\cite{root-05}.

\section{Numerical simulations}\label{section3}

In this section, we present numerical results for flow extraction based on multi-particle correlators and compare their performance with that of alternative approaches.
Our calculations reproduce results obtained in previous studies and further extend them to a broader context, thereby revealing additional subtleties in flow estimation from particle correlations in the presence of non-flow.

\subsection{Non-flow suppression in toy model I}\label{section3.1}

For toy model I, we note that there is an interplay between several competing factors:  
\begin{itemize}
\item Opening angle of the emitted particle pairs ($\phi_\mathrm{open}$): taking elliptic flow as an example, the additional emission of particle pairs at $90^\circ$ effectively suppresses the flow, whereas small or back-to-back opening angles do not significantly modify the input flow.

\item Order of the correlator: higher-order particle cumulants are expected to be less affected by the construction of particle pairs, since their contribution is primarily associated with two-particle correlations.

\item Correlation between the particle pair and the symmetry plane: particle pairs correlated with the symmetry plane lead to a smaller distortion of the input flow, although their role in particle correlators is more subtle than in the uncorrelated case.
\end{itemize}  

\begin{figure}[ht]
    \centering
    \begin{minipage}{0.4\textwidth}
        \centering
        \includegraphics[width=1.2\textwidth, height=0.3\textheight]{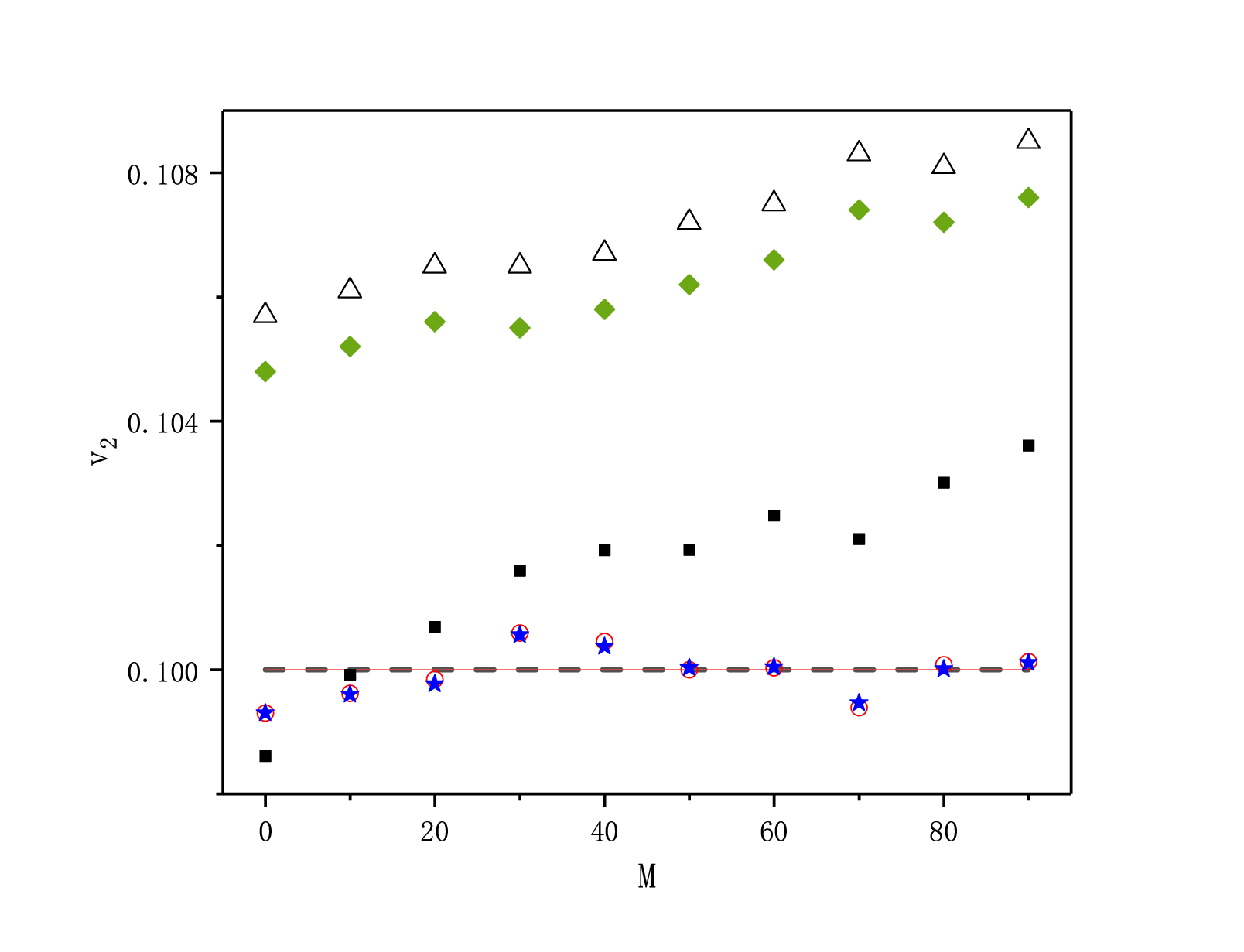}
    \end{minipage}
    \begin{minipage}{0.4\textwidth}
        \centering
        \includegraphics[width=1.2\textwidth, height=0.3\textheight]{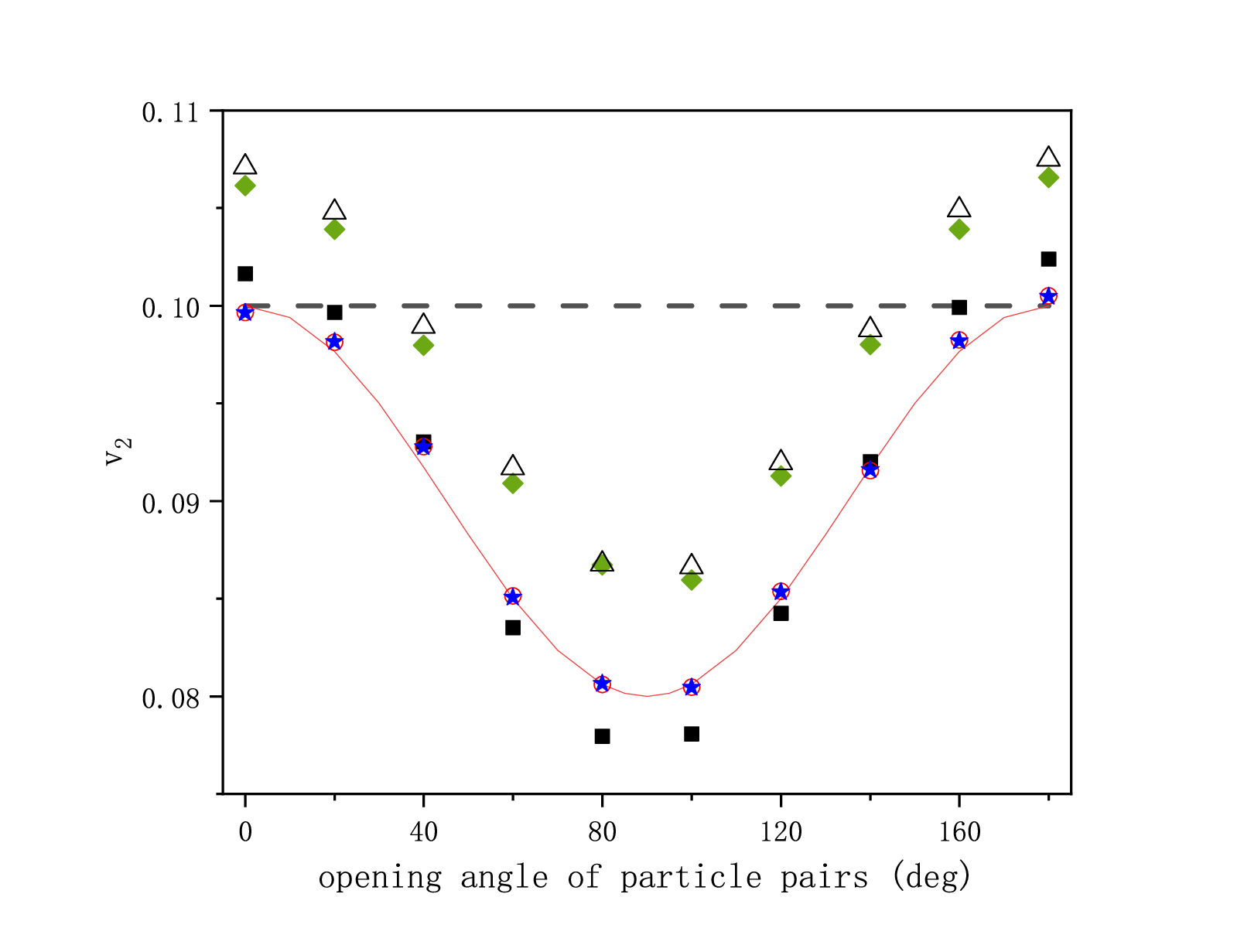}
    \end{minipage}
    \begin{minipage}{0.4\textwidth}
        \centering
        \includegraphics[width=1.2\textwidth, height=0.3\textheight]{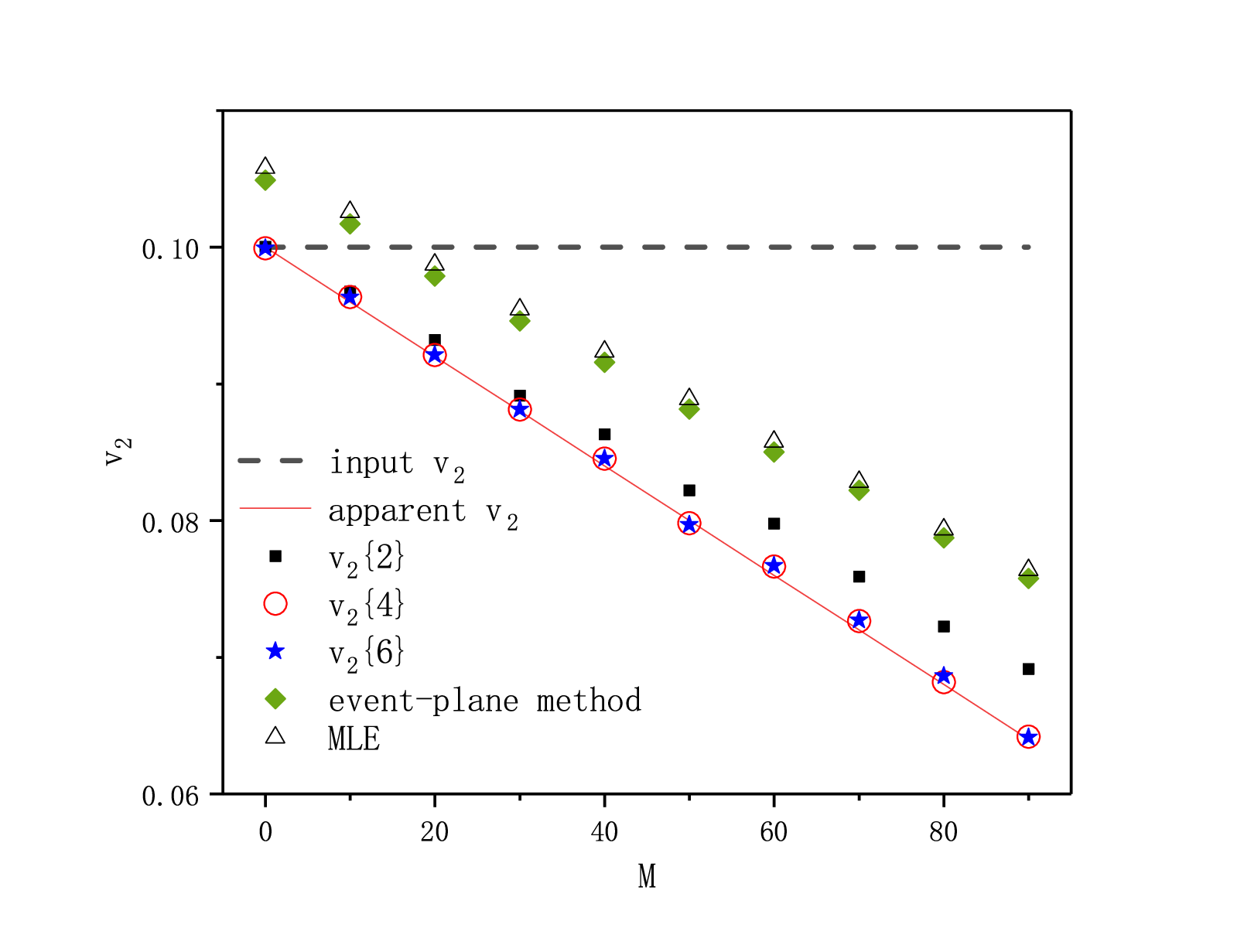}
    \end{minipage}
    \begin{minipage}{0.4\textwidth}
        \centering
        \includegraphics[width=1.2\textwidth, height=0.3\textheight]{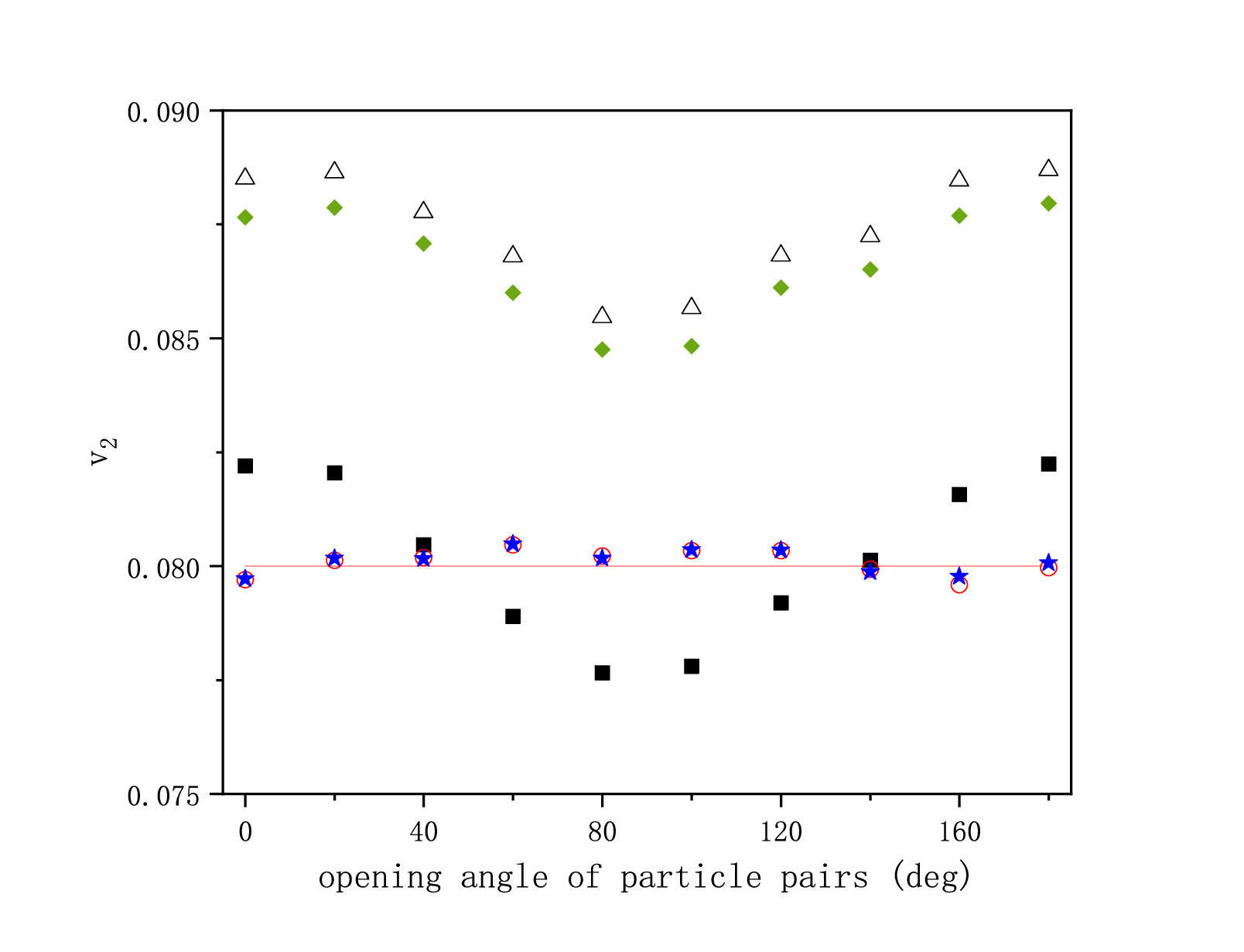}
    \end{minipage}
\renewcommand{\figurename}{Fig.}   
\caption{Elliptic flow $v_2$ evaluated using particle correlators, the event-plane method, and MLE.
The analysis is performed for 10,000 events, each of which contains a total of 500 particles.
The events are generated from the emission of particle pairs with a prescribed opening angle, superimposed on a background collective flow $v_2 = 0.1$ and $v_3=0.06$.
When applicable, the input elliptic flows are indicated by dashed black horizontal lines, the apparent values governed by Eqs.~\eqref{vnDecayUncorr} and~\eqref{vnDecayCorr} are shown by solid red curves.
Left column: The particle pairs are back-to-back, and the elliptic flow is evaluated as a function of the number of particle pairs.
Right column: The number of particle pairs is fixed to 50, and the elliptic flow is evaluated as a function of the opening angle.
Top row: The particle pairs are correlated with the symmetry plane, with one particle of each pair emitted according to the background one-particle distribution.
Bottom row: The same as the top row, but the particle pairs are uncorrelated with the symmetry plane.}
\label{v2_back_oa}
\end{figure}

In toy model I, for both the two-particle and four-particle correlators, the fraction of combinations effectively impacted by pair emission increases with the number of pairs.
However, the relative weight of such combinations is much smaller for the four-particle correlator.
In what follows, we refer to the Fourier coefficient evaluated according to
\begin{equation}
\tilde{v}_n = \langle \cos n(\phi - \Psi_n) \rangle
\label{vnAppDef}
\end{equation}
as the {\it apparent} flow harmonic~\footnote{In STAR's paper~\cite{RHIC-star-v2-07}, this is referred to as the {\it true} value.}, which typically differs from the {\it input} value owing to distortions induced by non-flow.
Nevertheless, from a theoretical perspective, one is often interested in extracting the underlying background flow masked by non-flow, which corresponds to the input flow harmonics in our numerical calculations.
In this regard, we will refer to the apparent value as {\it true} only when it coincides with the input flow harmonics.

\begin{figure}[ht]
    \centering
    \begin{minipage}{0.4\textwidth}
        \centering
        \includegraphics[width=1.2\textwidth, height=0.3\textheight]{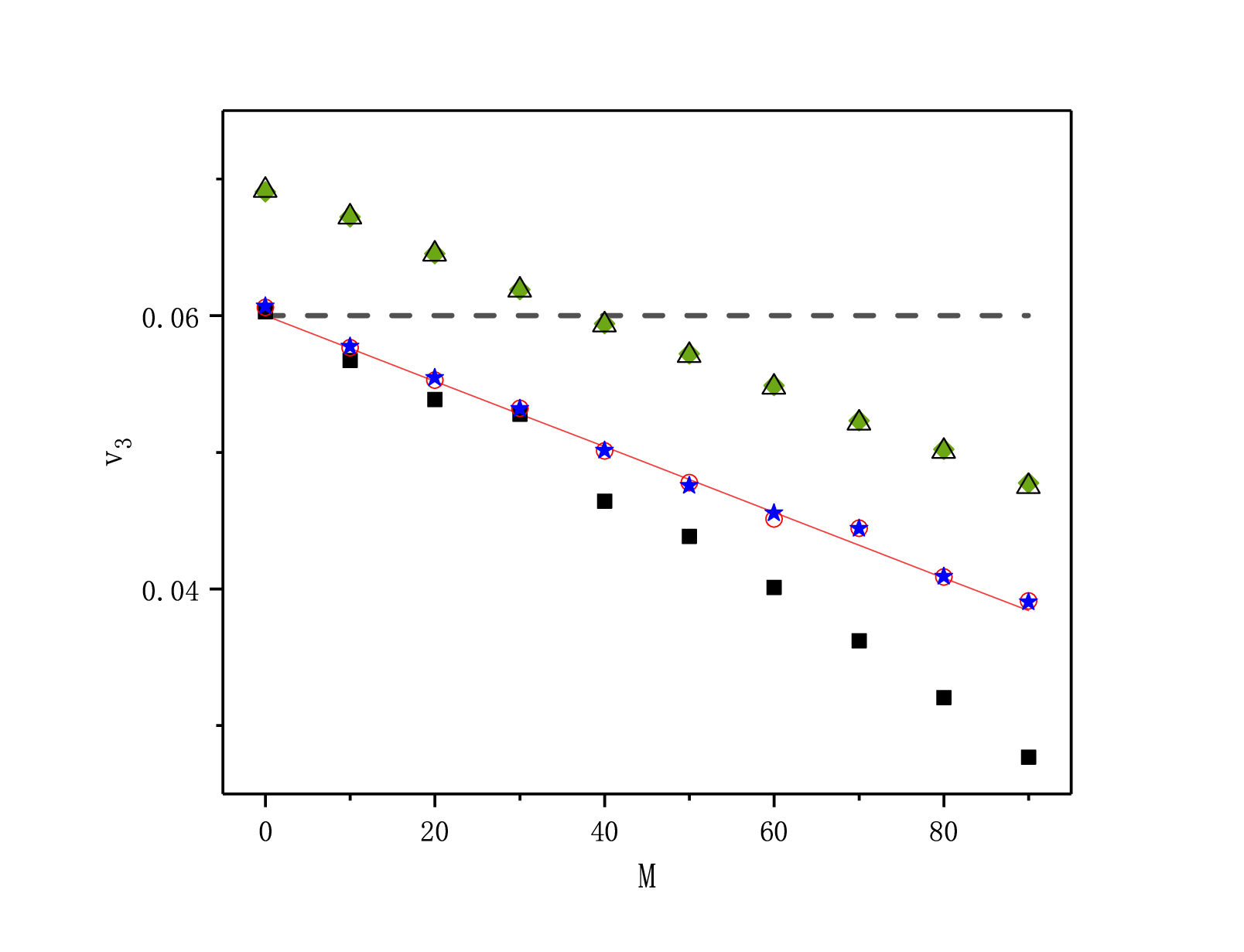}
    \end{minipage}
    \begin{minipage}{0.4\textwidth}
        \centering
        \includegraphics[width=1.2\textwidth, height=0.3\textheight]{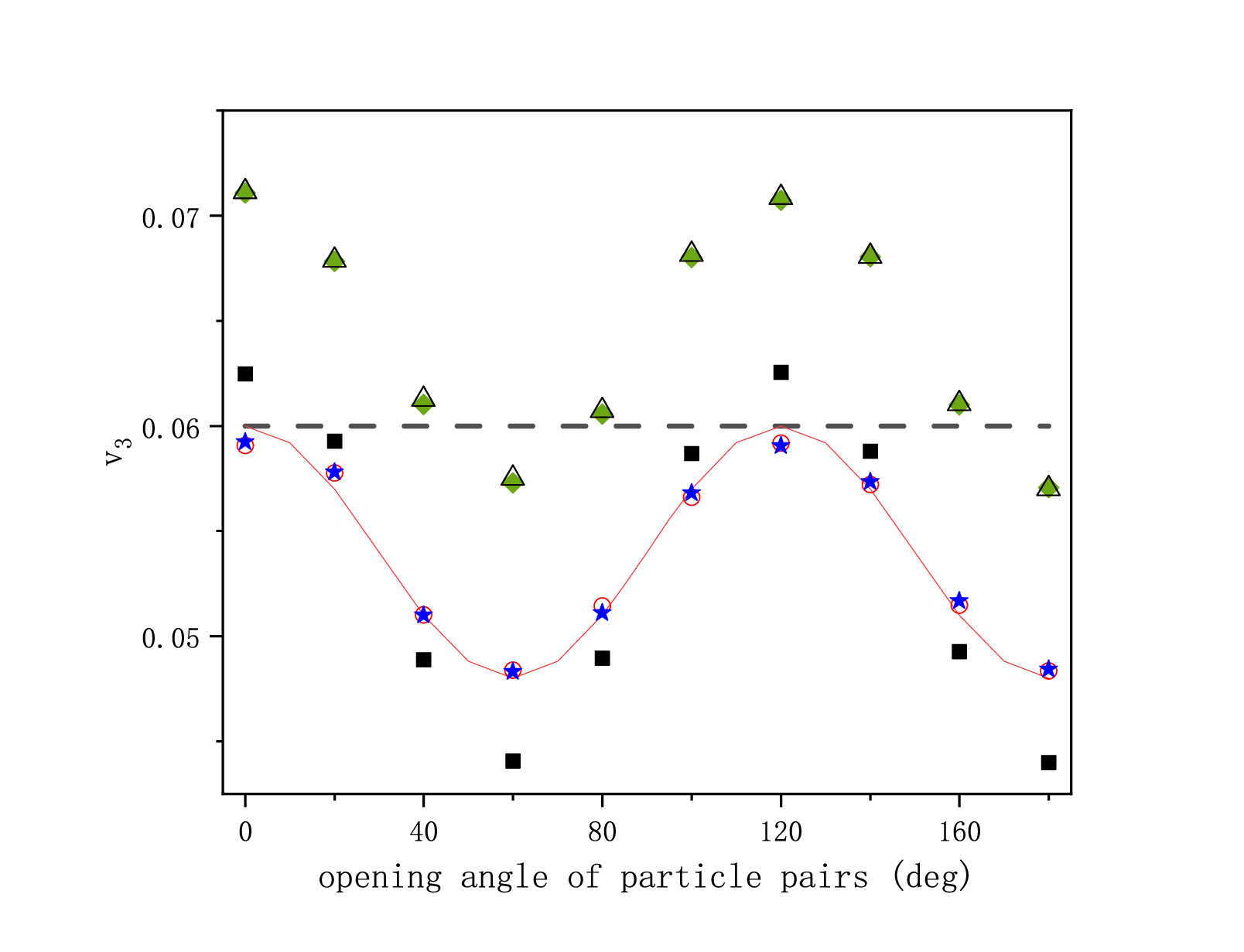}
    \end{minipage}
    \begin{minipage}{0.4\textwidth}
        \centering
        \includegraphics[width=1.2\textwidth, height=0.3\textheight]{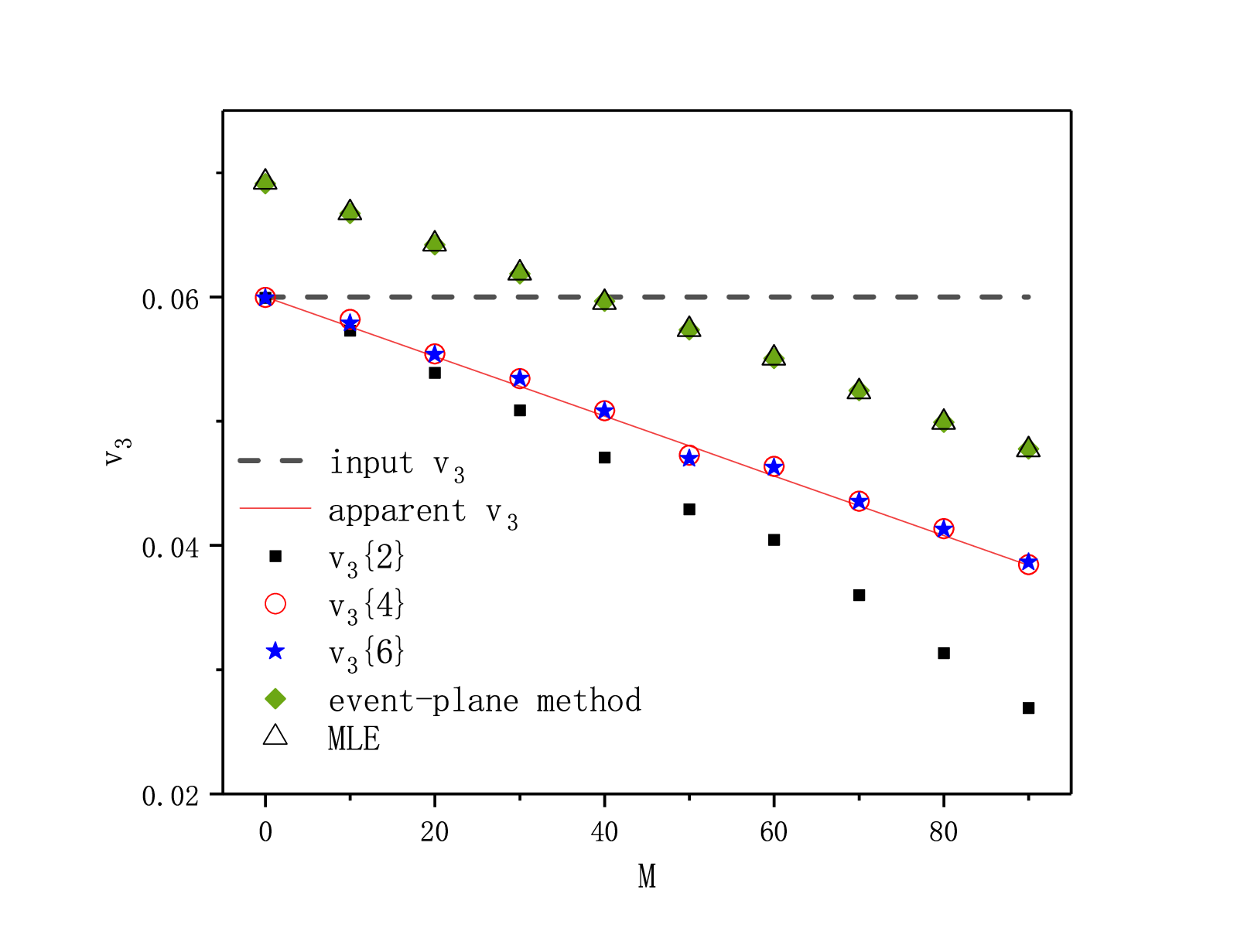}
    \end{minipage}
    \begin{minipage}{0.4\textwidth}
        \centering
        \includegraphics[width=1.2\textwidth, height=0.3\textheight]{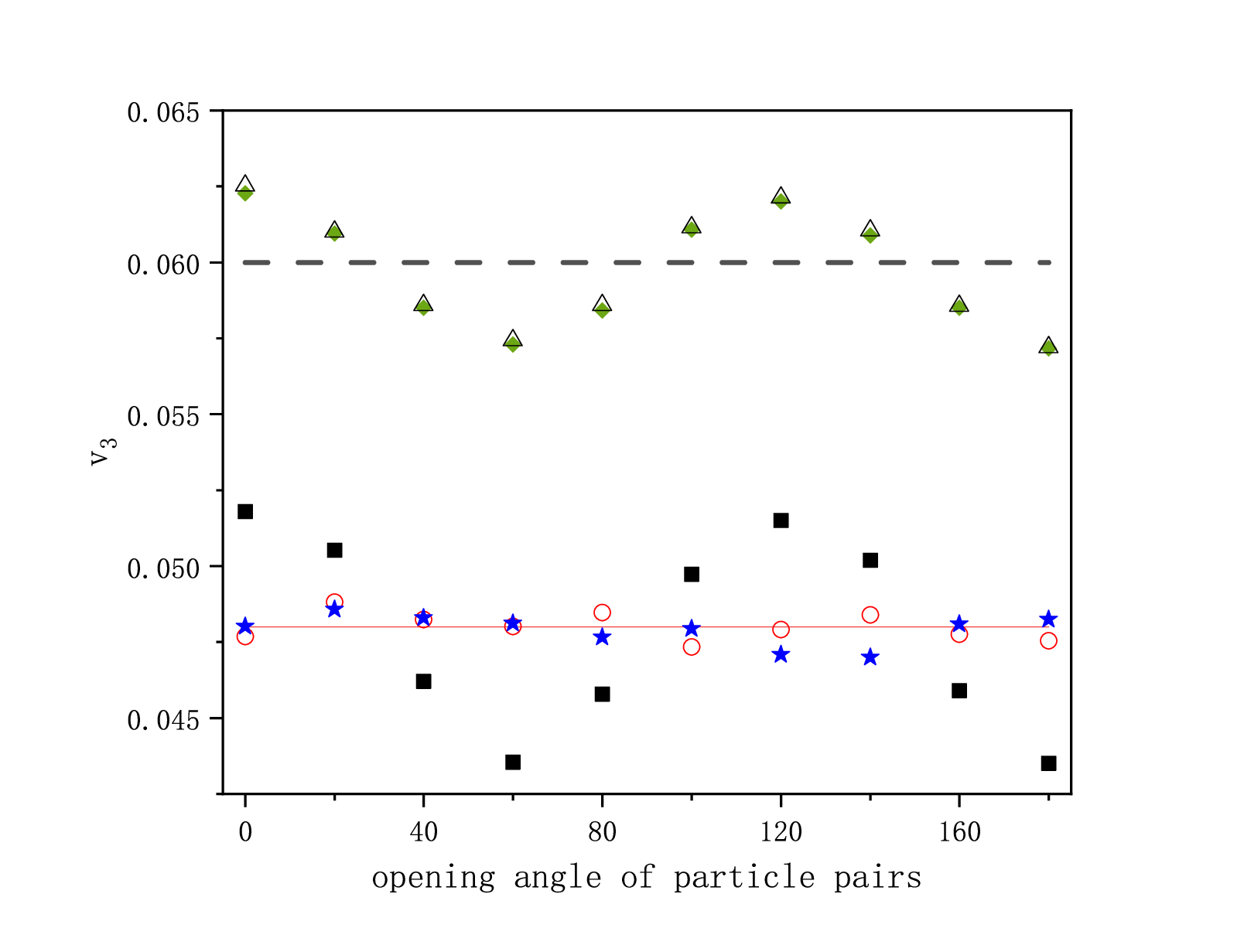}
    \end{minipage}
\renewcommand{\figurename}{Fig.}   
\caption{The same as Fig.~\ref{v2_back_oa}, but for the triangular flow $v_3$.}
\label{v3_back_oa}
\end{figure}

\begin{figure}[ht]
    \centering
    \begin{minipage}{0.4\textwidth}
        \centering
        \includegraphics[width=1.0\textwidth, height=0.25\textheight]{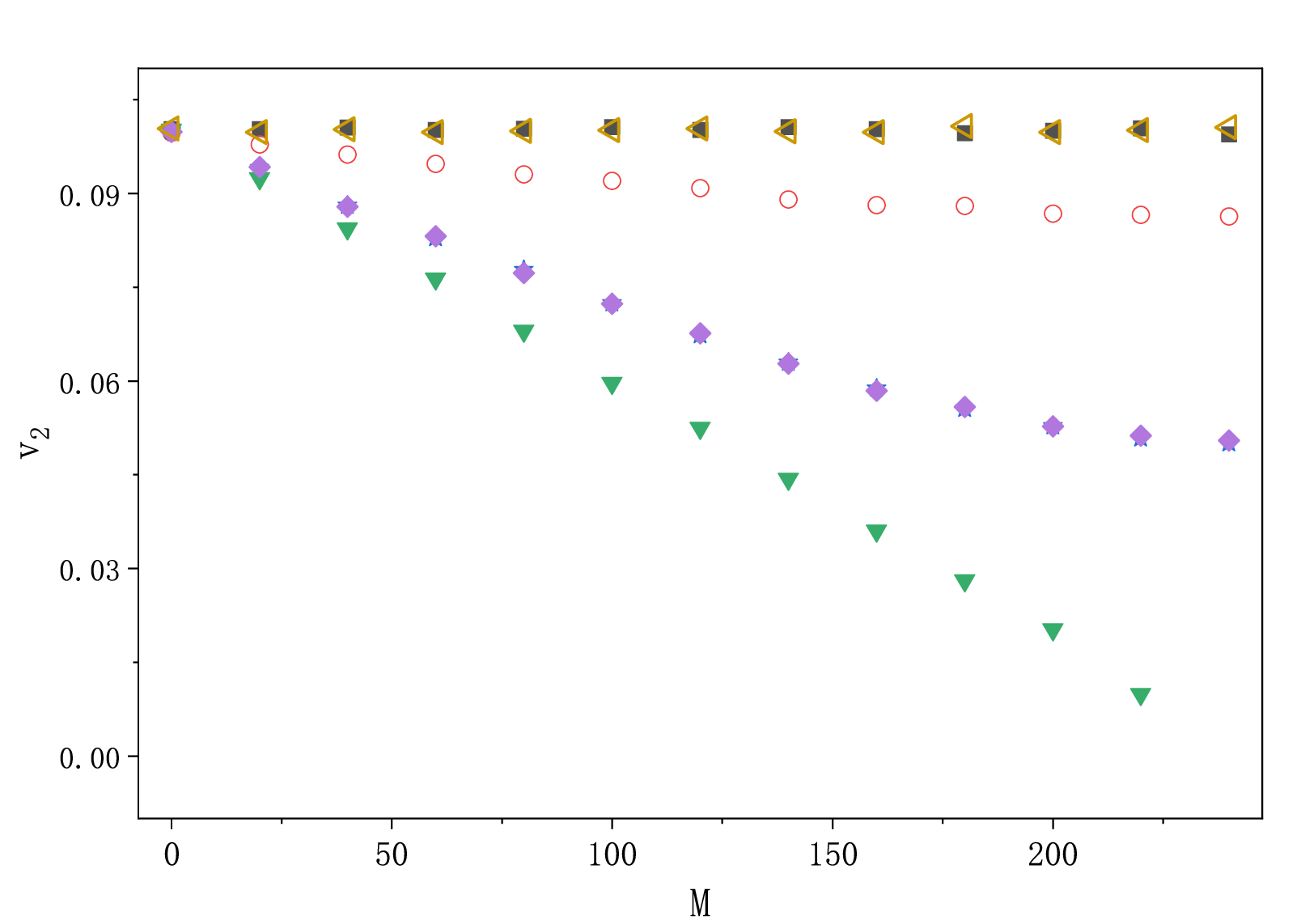}
    \end{minipage}
    \begin{minipage}{0.4\textwidth}
        \centering
        \includegraphics[width=1.0\textwidth, height=0.25\textheight]{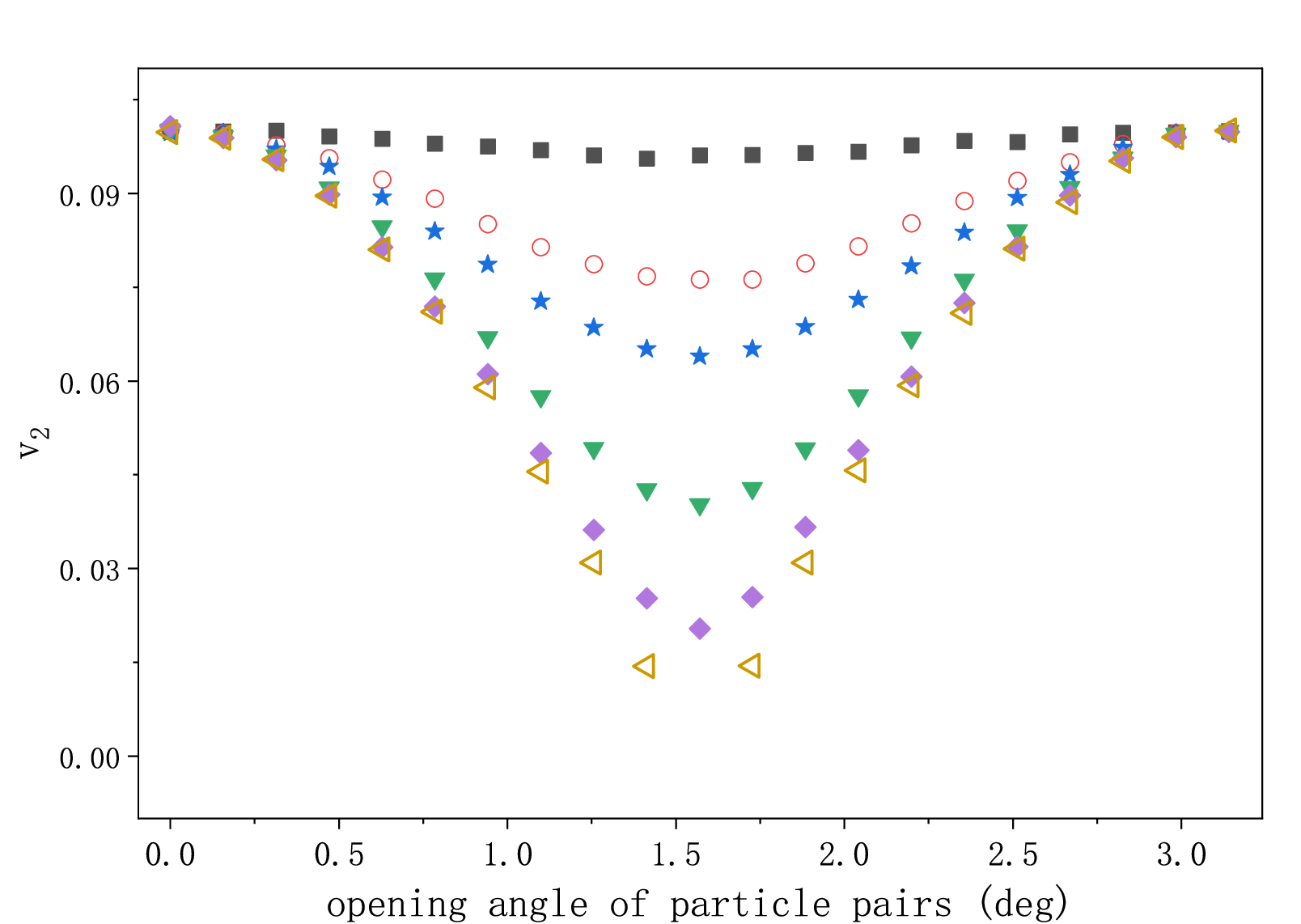}
    \end{minipage}
    \begin{minipage}{0.4\textwidth}
        \centering
        \includegraphics[width=1.0\textwidth, height=0.25\textheight]{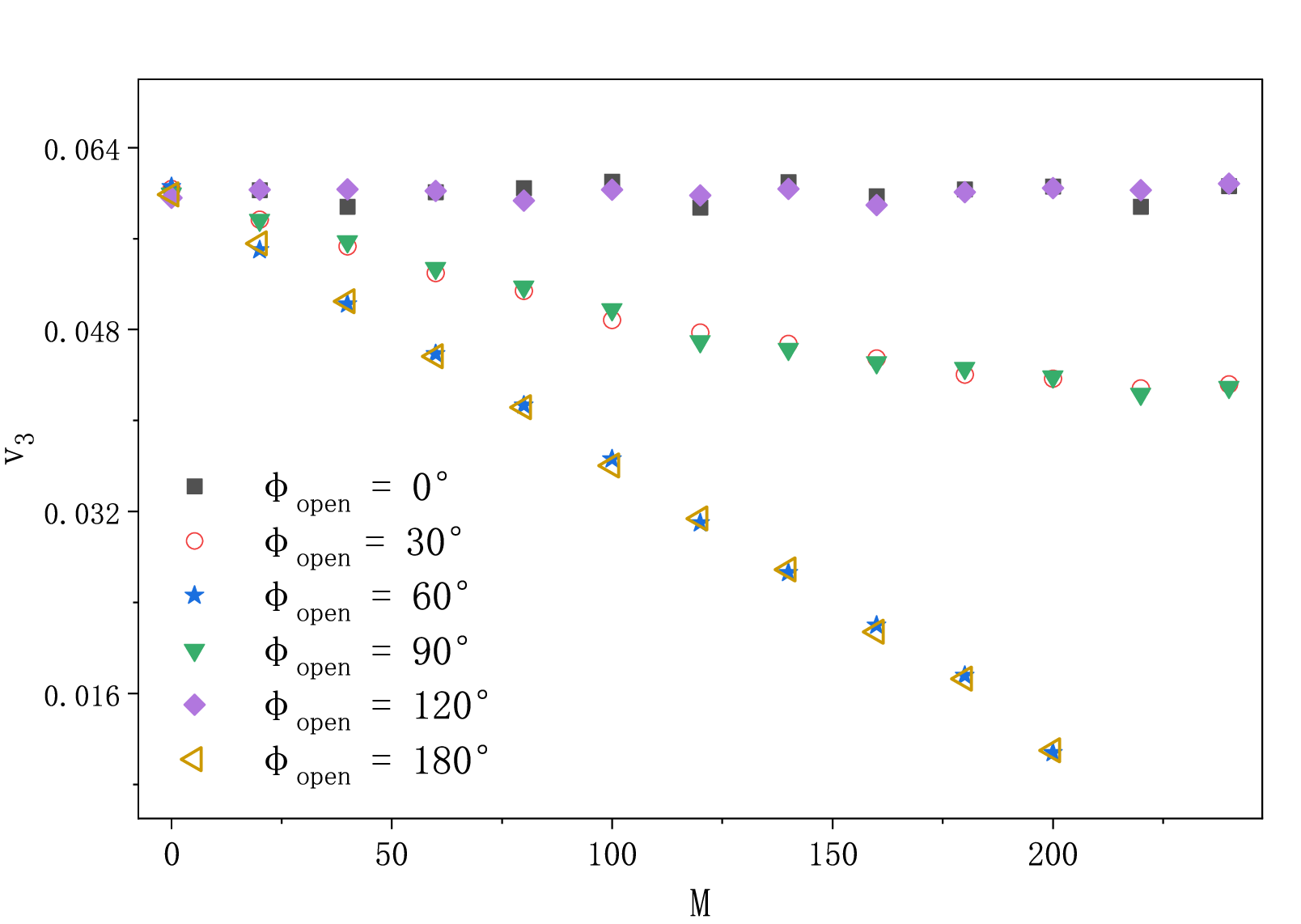}
    \end{minipage}
    \begin{minipage}{0.4\textwidth}
        \centering
        \includegraphics[width=1.0\textwidth, height=0.25\textheight]{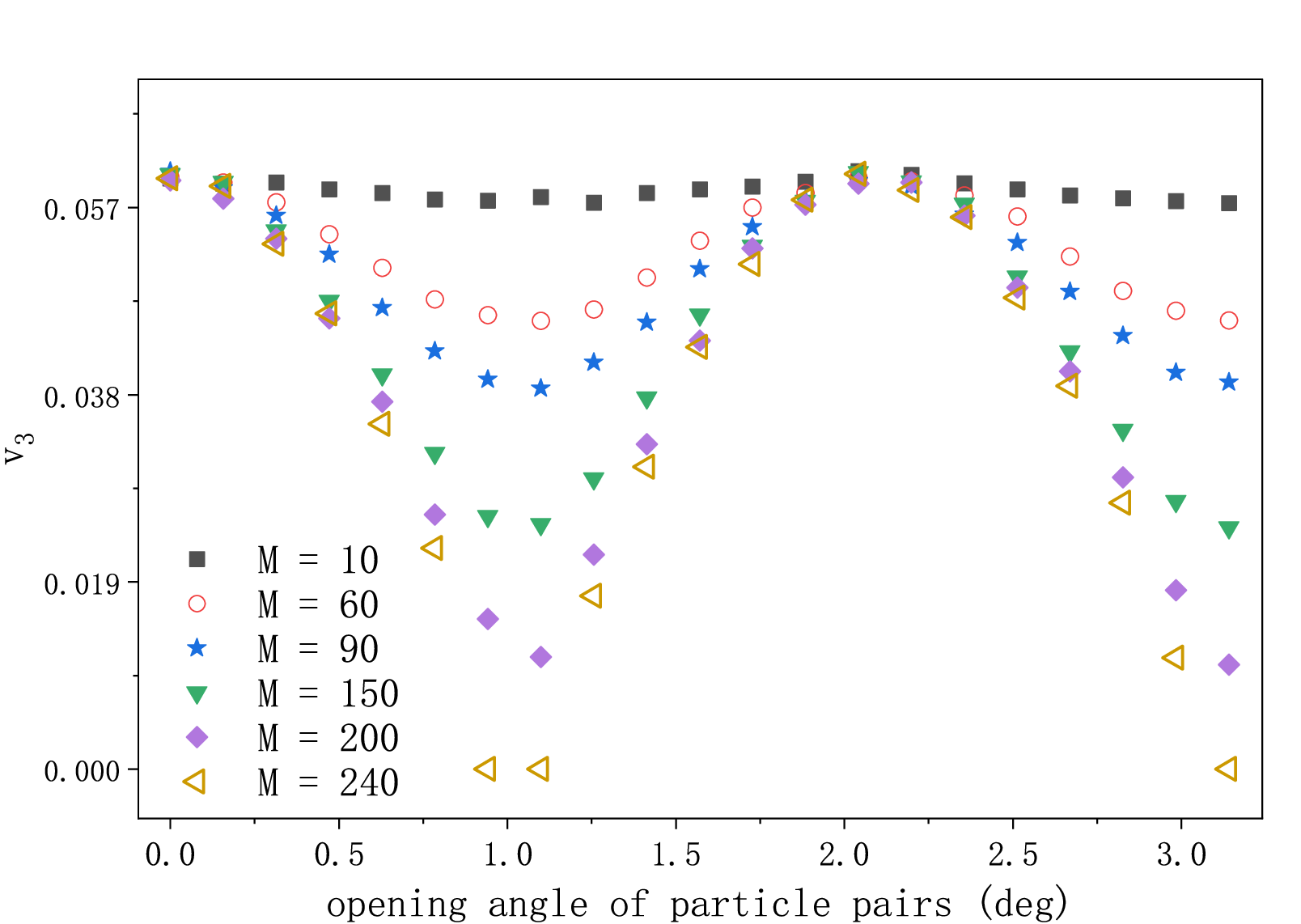}
    \end{minipage}
\renewcommand{\figurename}{Fig.}   
\caption{The elliptic and triangular flow harmonics using particle correlators $v_2\{4\}$ and $v_3\{4\}$.
The conventions are the same as Figs.~\ref{v2_back_oa} and~\ref{v3_back_oa}, but in the left column, we show the results for different opening angles $\phi_{\mathrm{open}}$, and in the right column, we present the results for different pair numbers $M$.
The calculations are carried out for the events where the particle pairs are correlated with the symmetry plane.}
\label{v23_variations}
\end{figure}

The numerical results are presented in Figs.~\ref{v2_back_oa},~\ref{v3_back_oa}, and~\ref{v23_variations}.
The analysis is performed on an event-by-event basis for 10{,}000 events, each of which contains a total of $500\ (=N+2M)$ particles.
Each event is generated from the emission of additional particle pairs with a prescribed opening angle, superimposed on a background collective flow with the harmonic coefficients $v_2 = 0.1$ and $v_3=0.06$.
When applicable, the input flow harmonics are indicated by a dashed black horizontal line, while the apparent values Eq.~\eqref{vnAppDef} are also indicated by a thin red line, using the analytical expressions, Eqs.~\eqref{vnDecayUncorr} and~\eqref{vnDecayCorr}.
These analytic expressions for the apparent values, along with those associated with the multi-particle correlators (cf. Eqs.~\eqref{vn2DecayUncorr},~\eqref{vn4DecayUncorr},~\eqref{vn2DecayCorr}, and~\eqref{vn4DecayCorr}), will be derived and further elaborated in the next section.

Fig.~\ref{v2_back_oa} shows the elliptic flow evaluated using the particle correlators $v_2\{2\}$, $v_2\{4\}$, and $v_2\{6\}$, compared with the results obtained from the event-plane method and the MLE.
We explore the dependence on the particle-pair number and opening angle for the two scenarios of toy model I.
First, we consider the emission of back-to-back particle pairs and compute the flow harmonics as functions of the number of pairs $M$.
These back-to-back pairs are taken to be either uncorrelated or correlated with the symmetry plane, as shown in the two panels in the left column.

In the top-left panel, we present $v_2\{2\}$, $v_2\{4\}$, and $v_2\{6\}$ as functions of the number of back-to-back pairs for events in which the pair emission is correlated with the symmetry plane.
In this case, the apparent flow harmonic coincides with the input value.
It is observed that the resulting $v_2\{4\}$ essentially agrees with $v_2\{6\}$, both slightly oscillating around the true value.  
In comparison, $v_2\{2\}$ overestimates the elliptic flow, and the deviation increases with the number of pairs.
We note that the behaviors of $v_2\{2\}$ and $v_2\{4\}$ are consistent with those reported in Ref.~\cite{RHIC-star-v2-07}.
In addition, we show the results obtained with the MLE and event-plane methods, which also exhibit an overestimation.
In this case, particle correlators perform consistently better than the other means.

In the bottom-left panel, we show the results for events in which the emission of particle pairs is uncorrelated with the symmetry plane.
In this case, the non-flow effect, quantified by the difference between the dashed black curve and the solid red one, becomes more pronounced compared to the case where the pair emission is correlated with the symmetry plane.
We observe that all three quantities $v_2\{2\}$, $v_2\{4\}$, and $v_2\{6\}$ align well with the apparent value of $v_2$ given by Eq.~\eqref{vnDecayUncorr}, which lies below the input value, with the deviation increasing as the pair number $M$ grows.
In particular, the results for $v_2\{4\}$ and $v_2\{6\}$ converge closely to the apparent value, while a visible deviation remains for $v_2\{2\}$, especially at larger $M$.
However, if the goal is to extract the genuine input flow harmonics, which in this case do not coincide with the apparent flow, one may conclude that all approaches deviate further from the input value as the number of particle pairs, and thus the impact of non-flow, increases.  

We analyze the dependence of the elliptic flow on the opening angle in the two panels in the right column of Fig.~\ref{v2_back_oa}, where the number of particle pairs is fixed to $M=50$.
Again, the additional pair emission that mimics particle decay is either correlated or uncorrelated with the background flow.
In the top-right panel, we show the results obtained for events in which the pair emission is correlated with the symmetry plane.
We find that for $v_2\{2\}$, the apparent elliptic flow is slightly overestimated at small and back-to-back angles but suppressed near $90^\circ$.
Although the differences between the elliptic flows estimated with different particle correlators are not large, $v_2\{4\}$ and $v_2\{6\}$ agree very well with the apparent values.
On the whole, as functions of the opening angle, all elliptic-flow estimators display a similar pattern characterized by a sizable suppression near $90^\circ$ and subsequently a larger deviation from the input value.
This suppression is naturally understood, since particles emitted out of the reaction plane tend to reduce the apparent elliptic flow, as given below in Eq.~\eqref{vnDecayCorr}.
When we compare the estimates obtained from particle correlators with those from the event-plane and MLE methods, we find that the latter exhibit a larger deviation from the apparent value than the particle-correlator estimates.
However, again, if the goal is to estimate the input flow, one might instead conclude that the event-plane and MLE methods provide a better estimate, owing to their closer agreement with the input value.
Nonetheless, upon closer inspection, the behavior of the event-plane method appears rather counter-intuitive.
In particular, if one chooses the symmetry plane to evaluate the flow harmonics, one should obtain the apparent value, which gives in fact the largest possible value of $v_2$ among different choices of $\Psi_2$.
How, then, can the event-plane method yield an estimate that exceeds the apparent value dictated by Eq.~\eqref{vnDecayCorr}?
The answer lies in event-by-event fluctuations: if one determines the event plane $\Psi_2$ separately for each event and then computes the elliptic flow, the resulting event-average indeed exceeds the apparent flow harmonic and approaches the input value.

In the bottom-right panel, we show the results for events in which the emission of particle pairs is uncorrelated with the symmetry plane. 
In this case, the apparent value Eq.~\eqref{vnDecayUncorr} becomes a constant that lies below the input value.
This behavior is intuitive: the non-flow contribution is isotropic by construction, and therefore the apparent flow signal is suppressed in a way that is independent of the details of the particle pair.
The particle correlators $v_2\{4\}$ and $v_2\{6\}$ provide an almost unbiased estimate of the apparent value, whereas $v_2\{2\}$ exhibits more pronounced oscillations.
We note that the behaviors of $v_2\{2\}$ and $v_2\{4\}$ are consistent with those shown in Fig.~11 of Ref.~\cite{RHIC-star-v2-07}, which has been used to illustrate the superior performance of multi-particle correlators as flow estimators.
Again, the event-plane method and the MLE provide estimates that exhibit the same modulation as $v_2\{2\}$, but remain closer to the input value.

In Fig.~\ref{v3_back_oa}, we show the results of the triangular flow $v_3$.  
The two panels in the top row correspond to events where the additional pair emissions are correlated with the symmetry plane, while those in the bottom row correspond to events where the pair emissions are uncorrelated with the symmetry plane.  
The two panels in the left column show the calculated triangular flow as a function of the number of particle pairs $M$, which are emitted in a back-to-back fashion.  
The two panels in the right column present the resulting triangular flow as a function of the opening angle $\phi_\mathrm{open}$, where the number of particle pairs is fixed at $M = 50$.

When compared with the results shown above in Fig.~\ref{v2_back_oa}, we observe the following distinct features.  
First, in the top-left panel of Fig.~\ref{v3_back_oa}, the apparent value is not constant but decreases linearly with increasing $M$. 
This behavior is a direct consequence of Eq.~\eqref{vnDecayCorr}, since $\cos(n\phi_\mathrm{open}) = -1$.  
Similar to the top-left panel of Fig.~\ref{v2_back_oa}, $v_3\{4\}$ essentially agrees with $v_3\{6\}$, exhibiting only small oscillations around the apparent value.  
In comparison, $v_3\{2\}$ underestimates the apparent triangular flow, and the deviation increases with the number of pairs.  
Although the specific value of the apparent flow governed by Eq.~\eqref{vnDecayUncorr} is quantitatively different, this feature largely persists in the case of uncorrelated pair emission shown in the bottom-left panel.

The two panels in the right column of Fig.~\ref{v3_back_oa} show the dependence on the opening angle $\phi_\mathrm{open}$.  
Again, for both cases, the higher-order correlators $v_3\{4\}$ and $v_3\{6\}$ stay closer to the apparent value compared to the lower-order one $v_3\{2\}$.  
When the pair emission is uncorrelated with the symmetry plane, due to its isotropic nature, the apparent flow is independent of the opening angle while remaining below the input value.  
The modulation of the apparent triangular flow shown in the top-right panel differs from that of the elliptic flow because the factor $\cos(n\phi_\mathrm{open})$ in Eq.~\eqref{vnDecayCorr} depends explicitly on the harmonic order $n$, leading to a different periodicity and different locations of maxima and minima for apparent $v_3$.

In Fig.~\ref{v23_variations}, we present the elliptic and triangular flow harmonics using particle correlators $v_2\{4\}$ and $v_3\{4\}$.  
We adopt essentially the same conventions as in Figs.~\ref{v2_back_oa} and~\ref{v3_back_oa}.  
In the left column, the dependence on the number of pairs is evaluated for different opening angles, while in the right column, different pair numbers are considered for flow harmonics as a function of the opening angle.  
The calculations are performed for events where the particle pairs are correlated with the symmetry plane.

From the two panels in the left column, for events where the pair emission is correlated with the symmetry plane, while the deviation from the input value increases with increasing pair number, one observes that the suppression is the most significant at $\phi_{\mathrm{open}}=90^\circ$ for the elliptic flow and $\phi_{\mathrm{open}}=60^\circ$ for the triangular flow.
The above feature is readily understood as Eq.~\eqref{vnDecayCorr} is not a function of $\phi_\mathrm{open}$, while the second term of Eq.~\eqref{vnDecayCorr} attains~$-1$ at $\phi_\mathrm{open} = \pi/n$.
From the two panels in the right column, one observes the same feature from a different perspective.
While the shape of the curves as functions of $\phi_\mathrm{open}$ remains unchanged, its distance from the input value increases with increasing $M$, which is intuitive as more additional particle pairs indicate a stronger non-flow effect.

The numerical results presented above indicate that the multi-particle correlators, particularly the higher-order ones, tend to approach the apparent flow.
This can be understood by noting that, by definition, higher-order cumulants naturally suppress lower-order correlations, as expected for several non-flow mechanisms, including particle decays.
However, if the non-flow also affects the overall collective flow, the apparent flow will systematically deviate from the input flow.
In the specific case of toy model I, flow estimates based on multi-particle correlators underestimate the input flow, whereas other estimators, such as event-plane methods, tend to overestimate the apparent flow and thereby yield values closer to the underlying genuine flow harmonics.
In the next section, using toy model II, we explore a scenario in which a multi-particle correlator picks up a high-order effective particle correlations, leading to a non-vanishing lower-order harmonic coefficient that is expected to vanish identically and does not appear in alternative approaches.

\subsection{Non-flow suppression in toy model II}\label{section3.2}

\begin{figure}[ht]
    \centering
    \begin{minipage}{0.4\textwidth}
        \centering
        \includegraphics[width=1.0\textwidth, height=0.25\textheight]{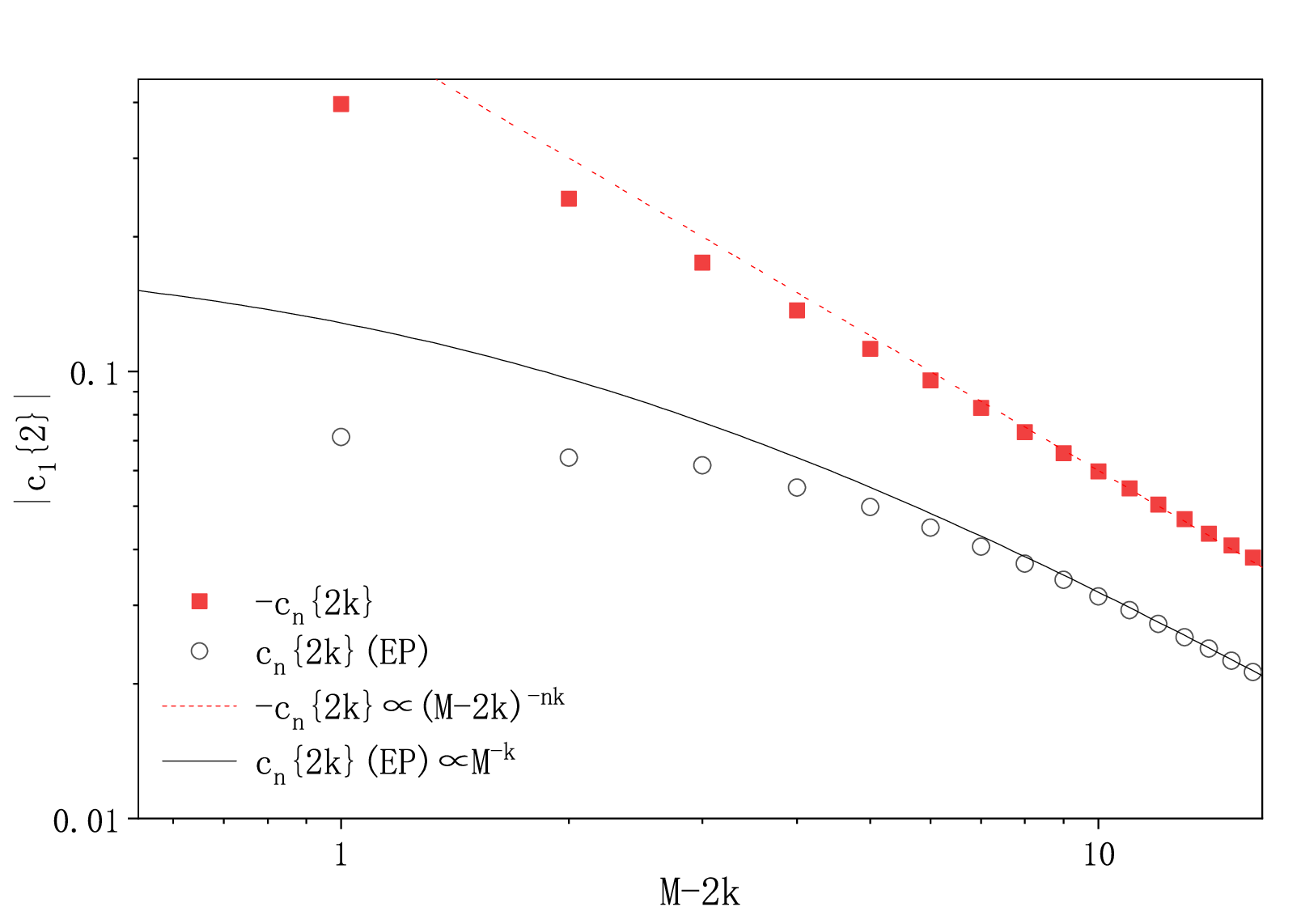}
    \end{minipage}
    \begin{minipage}{0.4\textwidth}
        \centering
        \includegraphics[width=1.0\textwidth, height=0.25\textheight]{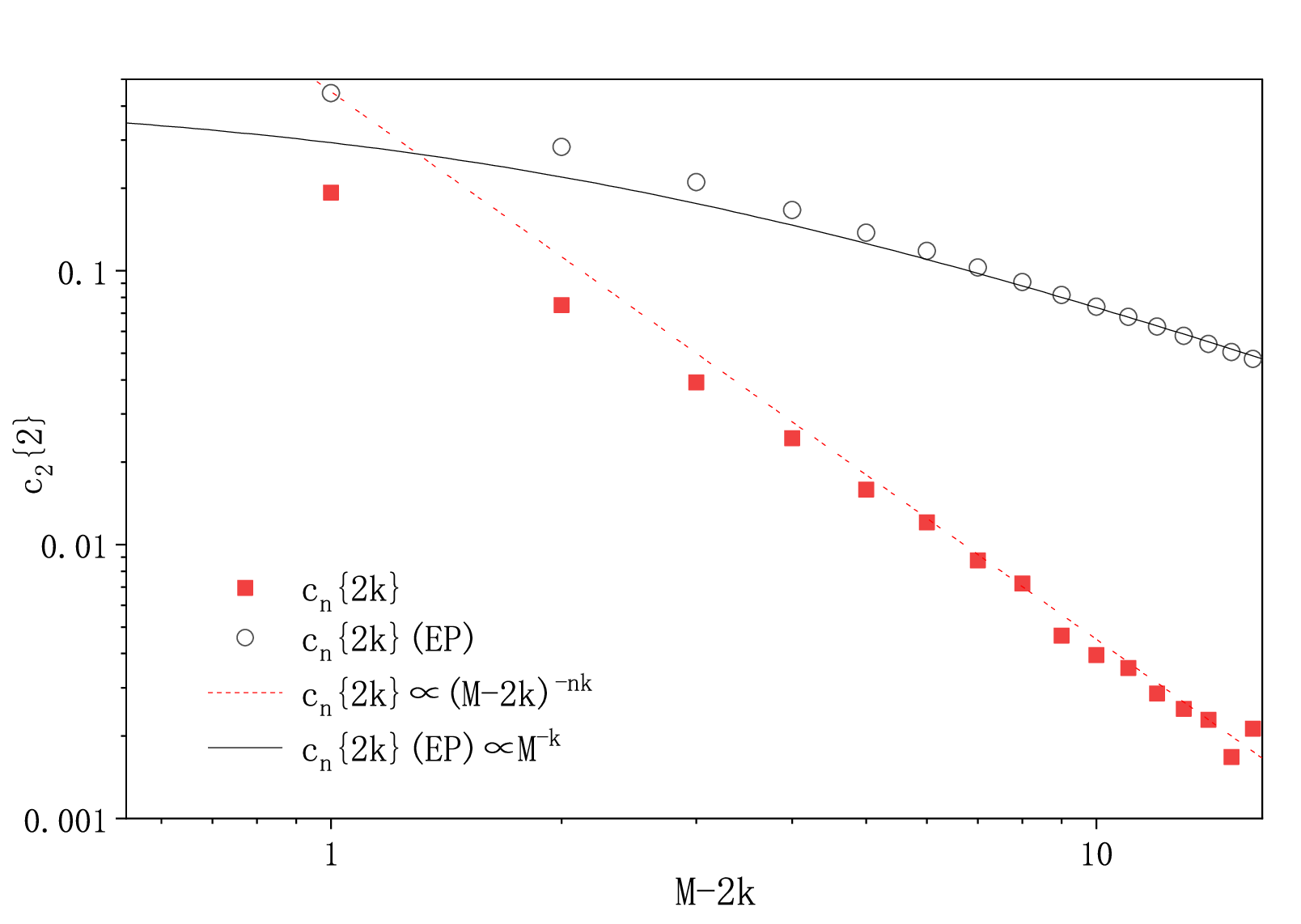}
    \end{minipage}
    \begin{minipage}{0.4\textwidth}
        \centering
        \includegraphics[width=1.0\textwidth, height=0.25\textheight]{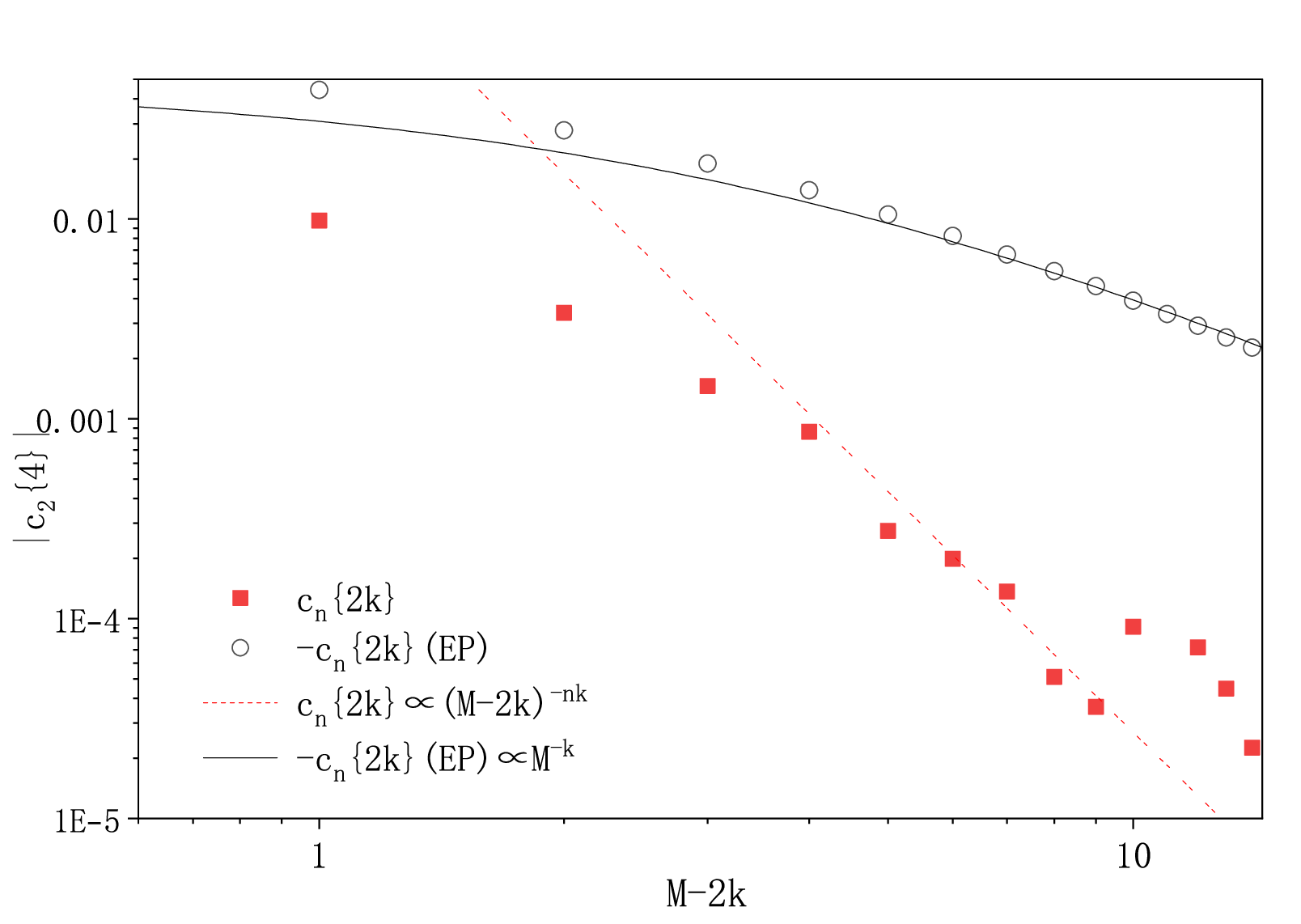}
    \end{minipage}
    \begin{minipage}{0.4\textwidth}
        \centering
        \includegraphics[width=1.0\textwidth, height=0.25\textheight]{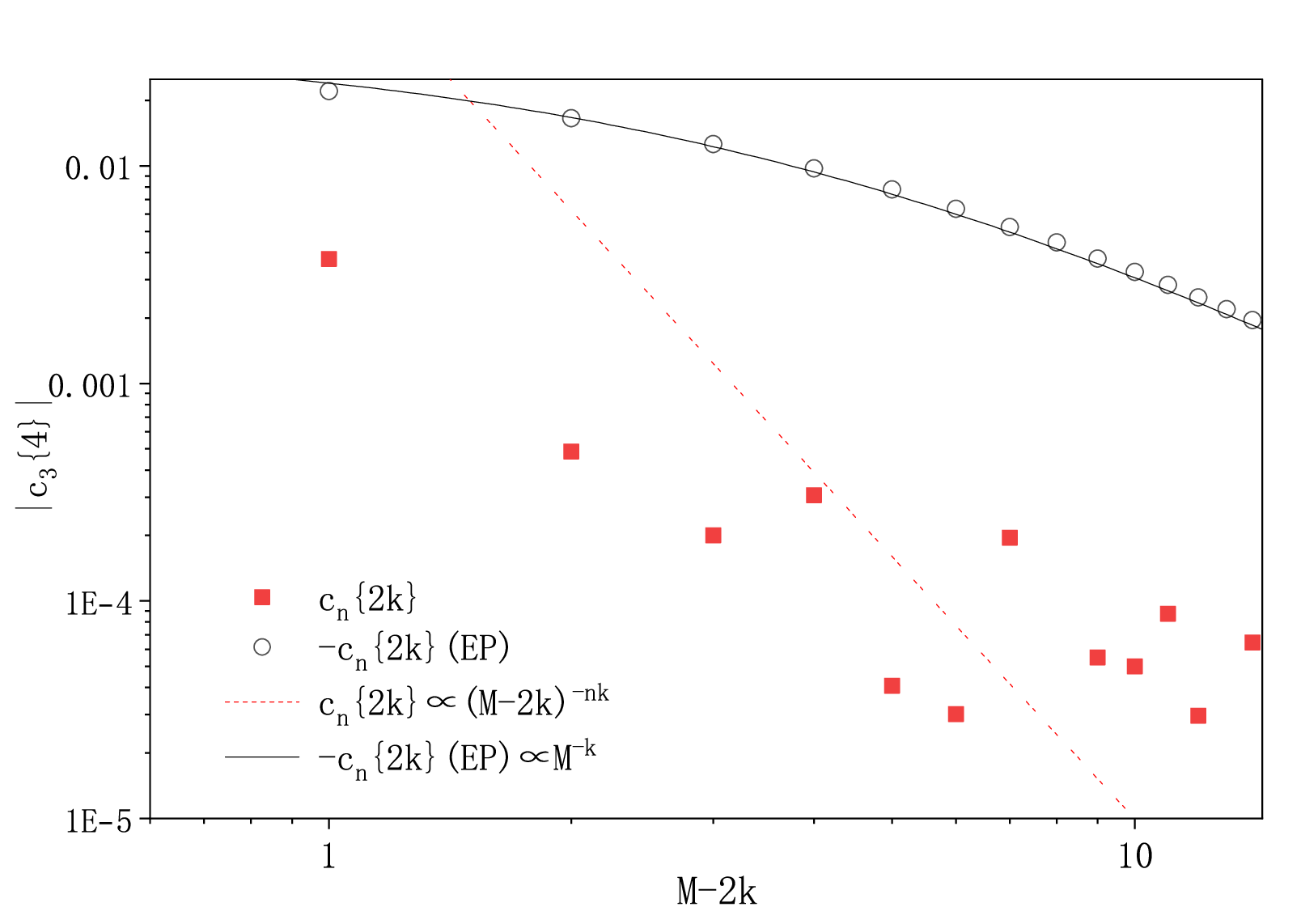}
    \end{minipage}
\renewcommand{\figurename}{Fig.}   
\caption{Multiplicity dependence of the cumulants $c_1\{2\}$, $c_2\{2\}$, $c_2\{4\}$, and $c_3\{4\}$ averaged over 100,000 random events without any background harmonic flow but with strict global momentum conservation, generated by the T-generation algorithm.
For comparison, the results are shown by filled red squares on a log-log scale and are compared with the fit $-nk \ln(M - 2k) + \mathrm{const}$, indicated by the red dashed lines.
They are also confronted with their counterparts, shown by black empty circles constructed from the flow harmonics obtained using the event-plane method.
The latter are compared against the fit $-k {\ln M} + \mathrm{const}$ (vs. $\ln(M-2k)$), indicated by the black solid curves.}
\label{cmp_scheme2}
\end{figure}

In this subsection, we present the numerical results for toy model II.
We consider events without any background harmonic flow, but with strict global momentum conservation, implemented within the same framework as the original study by Borghini, Dinh, and Ollitrault~\cite{hydro-corr-non-flow-01}.
The average $c_1\{2\}$, $c_2\{2\}$, $c_2\{4\}$, and $c_3\{4\}$ are evaluated using the cumulant definitions.
The analysis is performed on sets of 100,000 randomly generated events over a range of multiplicities, and the results are presented in Fig.~\ref{cmp_scheme2}.
In the top row of Fig.~\ref{cmp_scheme2}, we present $c_1\{2\}$ and $c_2\{2\}$ as functions of the multiplicity $M$.
The two panels in the bottom row show the four-particle cumulants $c_2\{4\}$ and $c_3\{4\}$.
For comparison, the results are shown on a logarithmic scale.
We observe that the results for $c_1\{2\}$~\cite{hydro-corr-non-flow-01} and $c_2\{2\}$~\cite{hydro-corr-non-flow-13} agree with the corresponding theoretical calculations.
Specifically, the negative sign of $c_1\{2\}$ can be derived analytically~\cite{hydro-corr-non-flow-01,hydro-corr-non-flow-02} and is a subtle consequence of the nonvanishing two-particle correlation induced by momentum conservation.
It was initially pointed out that the resulting directed flow $v_1\{2\}$ is the only nonvanishing collective flow at leading order, giving rise to a preferred azimuthal angle.
For higher-order harmonics, such as the elliptic flow, one can still obtain nonvanishing contributions by going to higher orders in the expansion of the normal distribution, which itself results from the central limit theorem.
It has been shown~\cite{hydro-corr-non-flow-13,hydro-corr-non-flow-20} that the magnitude of these quantities decreases with increasing multiplicity, following the scaling $c_n\{2k\} \propto 1/(M - 2k)^{nk}$~\cite{hydro-corr-non-flow-13,hydro-corr-non-flow-20}.
Consequently, $c_2\{2\}$ is considerably smaller in magnitude than $c_1\{2\}$, and the magnitudes of the four-particle cumulants are even smaller than those of the two-particle cumulants shown in the top row.

In this regard, toy model II provides a simple yet nontrivial example in which nonvanishing multi-particle cumulants arise purely from kinematic constraints, even in systems without any collective flow.
Consequently, flow estimates based on multi-particle correlators indicate nonzero flow harmonics.
From a physical point of view, this is somewhat puzzling.
To further investigate this issue, we compare these cumulants with their counterparts constructed from flow harmonics extracted using the event-plane method.
Specifically, under the assumption that the system is entirely governed by collective flow~\cite{hydro-corr-ph-04}, and by using the definition of cumulants together with Eq.~\eqref{oneParDis}, we have
\begin{eqnarray}
c_{1}\{2\} &=& \langle e^{i(\phi_1-\phi_2)}\rangle - \langle e^{i\phi_1}\rangle \langle e^{-i\phi_2}\rangle = v_{1}^{2} . \nonumber
\end{eqnarray}
Therefore, one can construct the corresponding cumulants from the flow harmonics extracted using the event-plane method by defining
\begin{eqnarray}
c_{1}\{2\}(\mathrm{EP}) &=& v_{1}^{2}(\mathrm{EP}) , \nonumber
\end{eqnarray}
along with analogous relations for higher harmonics given by
\begin{eqnarray}
c_{2}\{2\}(\mathrm{EP}) &=& v_{2}^{2}(\mathrm{EP}) , \nonumber \\
c_{2}\{4\}(\mathrm{EP}) &=& -\,v_{2}^{4}(\mathrm{EP}) , \nonumber \\
c_{3}\{4\}(\mathrm{EP}) &=& -\,v_{3}^{4}(\mathrm{EP}) . \nonumber
\end{eqnarray}
The results yield quantities with small magnitudes that are essentially different from those obtained via the definition of cumulants.
Moreover, numerical fits show that the cumulants evaluated from the flow harmonics obtained via the event-plane method asymptotically follow the scaling
\begin{equation}
c_n\{2k\} \propto \frac{1}{M^{k}} \, .\label{cnAssEV}
\end{equation}
Moreover, we note that $c_{1}\{2\}(\mathrm{EP})$ is positive, whereas $c_{2}\{4\}(\mathrm{EP})$ and $c_{3}\{4\}(\mathrm{EP})$ are negative, and these quantities have opposite signs compared to their counterparts evaluated using the definitions of cumulants.
We will elucidate this point further in the next section.

\section{An analytic account of the performance of multi-particle correlators}\label{section4}

In this section, we develop an analytic explanation for the behavior of the flow estimates based on multi-particle correlators discussed in the previous section.
These considerations are intended to clarify the main features of the cumulant-based estimates presented earlier.

\subsection{Toy model I}

In terms of the multi-particle correlators, the apparent flow is not evaluated according to Eq.~\eqref{vnAppDef} but via the cumulants.
Specifically, the definitions of $v_n\{2\}$ and $v_n\{4\}$ read~\cite{hydro-corr-ph-04, hydro-corr-ph-10}  
\bqn
v_n\{2\} &=& \sqrt{c_n\{2\}},\nb\\
v_n\{4\} &=& \sqrt[4]{-c_n\{4\}},\label{vn24Def}
\eqn  
where the two- and four-particle cumulants $c_n\{2\}$ and $c_n\{4\}$ are defined as  
\bqn
c_n\{2\} &=& \langle\langle 2 \rangle\rangle,\nb\\
c_n\{4\} &=& \langle\langle 4 \rangle\rangle - 2\,\langle\langle 2 \rangle\rangle^2 ,\label{cn24Def}
\eqn  
in terms of the event-averaged quantities $\langle\langle 2 \rangle\rangle$ and $\langle\langle 4 \rangle\rangle$, where the double brackets denote an average first over all particles in a given event and then over all events.  
Finally, the single-event two- and four-particle azimuthal correlations, $\langle 2 \rangle$ and $\langle 4 \rangle$, are defined as  
\bqn
\langle 2\rangle &=& \left\langle e^{in(\phi_1-\phi_2)} \right\rangle = \frac{1}{P(N, 2)} {\sum_{i,j}}^{\prime} e^{in(\phi_i-\phi_j)} , \nb\\
\langle 4\rangle &=& \langle e^{in(\phi_1+\phi_2-\phi_3-\phi_4)} \rangle = \frac{1}{P(N, 4)} {\sum_{i,j,k,l}}^{\prime} e^{in(\phi_i+\phi_j-\phi_k-\phi_l)} ,
\eqn  
where $P(N, M)=N!/(N-M)!$ is the permutation number and the prime indicates that all indices in the summation are distinct.  

If the particle emission is entirely governed by the flow, namely, individual particles are emitted independently according to the one-particle distribution function Eq.~\eqref{oneParDis}, it is readily verified $v_n\{2\}=v_n\{4\}=v_n$. 
The essence here is to evaluate the deviation from the pure flow owning to additional pair emission prescribed in toy model I.

We first consider the scenario where the additional emission is uncorrelated with the symmetry plane.
For a single event, the apparent flow harmonics, by Eq.~\eqref{vnAppDef}, now read
\bqn
\tilde{v}_n 
= \langle \cos n(\phi-\Psi_n)\rangle 
= v_n\frac{N}{N+2M} .
\label{vnDecayUncorr}
\eqn
Subsequently, the deviation from the input value $v_n$ takes the form
\bqn
\Delta v_n = \tilde{v}_n -v_n 
=-v_n \frac{2M}{N+2M} ,\label{vnDevUncorr}
\eqn
which indicates an overall underestimation of the input value, as observed from the event-plane and MLE methods.

Now, in terms of the two-particle cumulant, one needs to consider all different combination of particle pairs entering $\langle 2 \rangle$.
The additional emission of $M$ pairs of particles, mimicing particle decay, are carried out on top of $N$ particles entirely governed by the background flow.
The two-particle correlation involve the total permutation number $P(N+2M, 2)$, namely, the total number of ordered pairs from $(N+2M)$ particles, which can be written as a sum of
\bqn
P(N+2M, 2) = P(N, 2) + 2N(2M) + P(2M, 2) = P(N, 2) + 2N(2M) + 2M(2M-2) + 2M .
\eqn
Specifically, on the r.h.s. of the equality, there are $P(N, 2)$ pairs are entirely from the background flow, $2N(2M)$ pairs consisting one particle from the background flow and another from the particle decay, and $P(2M, 2)$ pairs are entirely from the particle decay, from which $2M(2M-2)$ are from different pairs and $2M$ are from the same pair.

As a result, the two-particle cumulant yields
\bqn
c_n\{2\} 
=\langle\langle 2\rangle\rangle
= v_n^2\frac{P(N, 2)}{P(N+2M, 2)}+ 0\frac{2N(2M)}{P(N+2M, 2)} + 0\frac{2M(2M-2)}{P(N+2M, 2)} + \cos(n\phi_\mathrm{open})\frac{2M}{P(N+2M, 2)} , \nb\\ \label{cn2DecayUncorr}
\eqn
where one notes that the contribution to $\langle\exp [in(\phi_1-\phi_2)]\rangle$ vanishes as long as the two particles forming the pair are uncorrelated.
Also, one notes that for the last term on the r.h.s. of the equality, one has 50\% chance to have a positive difference $\phi_1-\phi_2=+\phi_\mathrm{open}$ and 50\% chance to have $-\phi_\mathrm{open}$, and therefore the imaginary part of $\exp [in(\phi_1-\phi_2)]$ cancels out identically.
Subsequently, the flow harmonics $v_n\{2\}$ now read 
\bqn
v_n\{2\} &=& \sqrt{c_n\{2\}}  
=\left[v_n^2\frac{P(N, 2)}{P(N+2M, 2)}+ \cos(n\phi_\mathrm{open})\frac{2M}{P(N+2M, 2)}\right]^\frac12  \nb\\
&\simeq &  v_n \frac{N}{N+2M}+\frac{\cos(n\phi_\mathrm{open}){M}}{v_n N(N+2M)} ,\label{vn2DecayUncorr}
\eqn
where one has assumed that the second term is much smaller than the first one.
Therefore, the deviation from the input value $v_n$ takes the form
\bqn
\Delta v_n\{2\} = v_n\{2\} -v_n = -v_n\left(1-\sqrt{\frac{P(N, 2)}{P(N+2M, 2)}}\right) +\frac{\cos(n\phi_\mathrm{open}){M}}{v_n^2{P(N, 2)}} .\label{vn2DevUncorr}
\eqn
One sees immediately that the first term on the r.h.s. of the equality gives an overall underestimation of the input value.
Comparing Eq.~\eqref{vnDevUncorr} with Eq.~\eqref{vn2DevUncorr}, as $\sqrt{\frac{N(N-1)}{(N+2M)(N+2M-1)}}\lesssim \frac{N}{N+2M}$, the overall underestimation is similar.
However, the numerical calculations presented earlier indicate that both cases are worse than the event-plane and MLE estimations.
Besides, the factor $\cos(n\phi_\mathrm{open})$ of the second term provide an additional oscillation with a period in $\phi_\mathrm{open}$ of a form depending on the harmonic order $n$
\bqn
T= \frac{2\pi}{n} , \label{TOscillationPeriod}
\eqn
which is a unique feature associated with particle correlators.
It is worth noting that the approximation on the second line of Eq.~\eqref{vn2DecayUncorr} is not always valid.
As $1/v_n^2$ is potentially a big number, the second term might be sizable compared with the first term when $\phi_\mathrm{open}$ is a multiple of $\pi/n$.
However, this does not affect the periodic nature given by Eq.~\eqref{TOscillationPeriod}.
This feature has been manifestly shown by numerical simulations in the Sec.~\ref{section3}.

To discuss the flow harmonics evaluated in terms of the four-particle cumulant, we again perform the counting of combinatorial number first.
We reiterate the total permutation number $P(N+2M, 4)$ into the sum
\bqn
P(N+2M, 4) &=& P(N, 4) + 4 P(N, 3)P(2M, 1) + 6 P(N,2)P(2M,2) + 4 P(N, 1)P(2M, 3) + P(2M, 4),\label{PN+2M4}
\eqn
where each term on the r.h.s. of the equality corresponds to different ways of splitting the four particles into background flow and additional pair emission.
For each case, it will divide into different scenarios.
Fortunately, the calculation is relatively straightforward when the pair emission is uncorrelated to the symmetry plane.
We have
\bqn
\langle\langle 4 \rangle\rangle &=& v_n^4\frac{P(N, 4)}{P(N+2M, 4)} + 0\frac{4 P(N, 1)P(2M, 3)}{P(N+2M, 4)} + v_n^2\cos (n\phi_\mathrm{open})\frac23\frac{6 P(N,2)2M}{P(N+2M, 4)} \nb\\
&&+ 0\frac{4 P(N, 3)P(2M, 1)}{P(N+2M, 4)} + \cos^2 (n\phi_\mathrm{open})\frac23\times\frac{4!}{2!}\frac{P(M, 2)}{P(N+2M, 4)} ,\label{bra4Uncorr}
\eqn
where we note that two thirds of all $4!$ possible permutations of four particles furnished by two ordered pairs from the additional particles emission will contribute nontrivially to $\langle\exp[{i(\phi_1+\phi_2-\phi_3-\phi_4)}]\rangle$ and, therefore, the relevant combinatorial number for four particles from two background particles and one emission pair to form two pairs is $8 P(N,2) M$ and that from two emission pairs is $ 8 P(M, 2)$.
The result of Eq.~\eqref{bra4Uncorr} can be substituted into Eq.~\eqref{cn24Def} and then Eq.~\eqref{vn24Def} to evaluate the flow $v_n\{4\}$.
One observes that the oscillation originated from the factor $\cos (n\phi_\mathrm{open})$ is still there, but its impact is suppressed due to an almost perfect cancellation between the corresponding terms of Eq.~\eqref{bra4Uncorr} and those appear in $2\langle\langle 2\rangle\rangle^2$ given by Eq.~\eqref{cn2DecayUncorr}.
Specifically, one has
\bqn
v_n^2\cos (n\phi_\mathrm{open})\frac23\frac{6 P(N,2)2M}{P(N+2M, 4)} - 2\times 2v_n^2\frac{P(N, 2)}{P(N+2M, 2)} \cos(n\phi_\mathrm{open})\frac{2M}{P(N+2M, 2)} \sim 
\frac{32\,M\,N^2}{(N+2M)^5}v_n^2\cos\left(n\phi_{\mathrm{open}}\right)  \label{appEx1}
\nb \\ 
\eqn
and
\bqn
\cos^2 (n\phi_\mathrm{open})\frac23\times\frac12\frac{4!P(M, 2)}{P(N+2M, 4)} - 2\times \left[\cos(n\phi_\mathrm{open})\frac{2M}{P(N+2M, 2)}\right]^2 \sim \frac{32\,M^2}{(N+2M)^5}\,
\cos^2\left(n\phi_{\mathrm{open}}\right) , \label{appEx2} 
\eqn
both of which are suppressed roughly by a factor $1/(N+2M)$.
Moreover, in particular, one notes that $\cos^2 (n\phi_\mathrm{open})= \frac12\left(1+\cos (2n\phi_\mathrm{open})\right)$, therefore the residue consists of an oscillation with half of the period given by Eq.~\eqref{TOscillationPeriod}, which is also observed in the numerical simulations.
At first glance, the cancellation observed in $v_n\{4\}$ may appear ``accidental.''  
However, as we will see again in the case of correlated pair emission, such a cancellation arises once more.  
Therefore, as elaborated further below, the terms that cancel out are understood to originate primarily from two-particle correlations, which are, to a significant extent, designed to be canceled out by construction.  
The resulting flow harmonics are given by  
\bqn
v_n\{4\} &=& \sqrt[4]{  2\,\langle\langle 2 \rangle\rangle^2 - \langle\langle 4 \rangle\rangle}\nb\\
&\simeq&\sqrt[4]{v_n^4\frac{P(N, 2)P(N, 2)}{P(N+2M, 2)P(N+2M, 2)}+v_n^4\frac{8\,M\,N^3}{(N+2M)^5}-\frac{32\,M\,N^2}{(N+2M)^5}v_n^2\cos\left(n\phi_{\mathrm{open}}\right)-\frac{32\,M^2}{(N+2M)^5}\,\cos^2\left(n\phi_{\mathrm{open}}\right)} \nb\\
&\simeq& v_n\frac{N}{N+2M}+\frac14\left[\frac{v_n8M}{(N+2M)^2}-\frac{32M}{(N+2M)^2N}\frac{\cos\left(n\phi_{\mathrm{open}}\right)}{v_n}-\frac{32M^2}{(N+2M)^2N^3}\frac{\cos^2\left(n\phi_{\mathrm{open}}\right)}{v_n^3}\right].\nb\\
\label{vn4DecayUncorr}
\eqn

Secondly, if the additional emission is correlated to the symmetry plane, the calculations follow the same strategy but become more tedious.
First, Eq.~\eqref{vnAppDef} gives
\bqn
\tilde{v}_n 
= \langle \cos n(\phi-\Psi_n)\rangle 
= v_n\left[\frac{N+M}{N+2M}+ \cos (n\phi_\mathrm{open})\frac{M}{N+2M}\right],  \label{vnDecayCorr}
\eqn
where one notes when the particle from the additionally emitted pair is not aligned with the symmetry plane $\Psi_n$, it involves the integral
\bqn
\frac{1}{2\pi}\int d\phi \cos n(\phi-\Psi_n) 2v_n\cos n(\phi+\phi_\mathrm{open}-\Psi_n) 
= v_n\cos (n\phi_\mathrm{open}) .
\eqn
On the one hand, Eq.~\eqref{vnDecayCorr} coincides with the input value when $\phi_\mathrm{open}=2\pi/n$, as in the case of $v_2$ with perfectly aligned and back-to-back pair emission shown in the left column of Fig.~\ref{v2_back_oa}.
On the other hand, when $\phi_\mathrm{open}=\pi/n$, deviates the most from the input value as discussed in Fig.~\ref{v23_variations}.
Thus if ones goal is to extract the flow harmonics of the background flow, the latter corresponds to the most biased estimation for the flow.

For the two-particle cumulant, Eq.~\eqref{cn2DecayUncorr} becomes
\bqn
c_n\{2\} 
&=& \langle\langle 2\rangle\rangle \nb\\
&=& v_n^2\frac{P(N, 2)}{P(N+2M, 2)}
+ \frac12\left(1+\cos (n\phi_\mathrm{open})\right)v_n^2\frac{2N(2M)}{P(N+2M, 2)} \nb\\
&&+ \frac12\left(1+\cos (n\phi_\mathrm{open})\right)v_n^2\frac{2M(2M-2)}{P(N+2M, 2)}
+ \cos(n\phi_\mathrm{open})\frac{2M}{P(N+2M, 2)} \nb\\
&=& v_n^2\frac{ P(N+M,2)+P(M,2)}{P(N+2M,2)}
+\frac{2M}{P(N+2M,2)}\left[(N+M-1)v_n^2 + 1\right]\cos\left(n\phi_{\mathrm{open}}\right) ,\label{cn2DecayCorr}
\eqn
where one notices that two cross terms are now non-vanishing, and one uses
\bqn
\frac{1}{(2\pi)^2}\int d\phi_1 d\phi_2  (2v_n)^2\cos n(\phi_1-\Psi_n) \cos n(\phi_2+\phi_\mathrm{open}-\Psi_n) \cos n(\phi_1-\phi_2) = v_n^2 \cos (n\phi_\mathrm{open}) ,\nb
\eqn
and
\bqn
\frac{1}{(2\pi)^2}\int d\phi_1 d\phi_2  (2v_n)^2\cos n(\phi_1+\phi_\mathrm{open}-\Psi_n) \cos n(\phi_2+\phi_\mathrm{open}-\Psi_n) \cos n(\phi_1-\phi_2) = v_n^2 .\nb
\eqn
The last term in Eq.~\eqref{cn2DecayCorr} arises solely from the correlation between the two particles forming a pair and is therefore unrelated to the background flow, which is expected to be eliminated in higher-order cumulants.
Subsequently, the flow harmonics $v_n\{2\}$ now read
\bqn
v_n\{2\} &=&  
v_n\left\{\frac{ P(N+M,2)+P(M,2)}{P(N+2M,2)}+\frac{2M}{P(N+2M,2)}\left[(N+M-1) + \frac{1}{v_n^2}\right]\cos\left(n\phi_{\mathrm{open}}\right)
\right\}^\frac12 \nb\\
&\simeq& v_n\sqrt{\frac{P(N+M,2)+P(M,2)}{P(N+2M, 2)}}
\left[1+ \cos (n\phi_\mathrm{open})\frac{M}{P(N+M,2)+P(M,2)}\left[(N+M-1) + \frac{1}{v_n^2}\right]
\right]   \nb\\
&\simeq& v_n\frac{\sqrt{(N+M)^2+M^2}}{N+2M}
\left[1+ \cos (n\phi_\mathrm{open})\frac{M}{P(N+M,2)+P(M,2)}\left[(N+M-1) + \frac{1}{v_n^2}\right]
\right] .\label{vn2DecayCorr}
\eqn
The last line can be compared with the apparent value in Eq.~\eqref{vnDecayCorr}.  
One finds that the coefficients of the modulation $\cos\left(n\phi_{\mathrm{open}}\right)$ of the two expressions are essentially identical up to a shift of $\frac{1}{v_n^2}\frac{M}{P(N+M,2)}$.  
The shift $\frac{1}{v_n^2}\frac{M}{P(N+M,2)}$ can be readily used to account for the overshooting of $v_n\{2\}$ at $\phi_\mathrm{open} \sim 0$ and $\pi$, and the undershooting at $\phi_\mathrm{open} \sim \pi/2$.

By comparing Eq.~\eqref{vn2DecayCorr} with the uncorrelated case, Eq.~\eqref{vn2DecayUncorr}, one finds that the overall estimation becomes less severe, as intuitively expected and observed in numerical calculations.
The same type of oscillation, modulated by the factor $\cos (n\phi_\mathrm{open})$, is observed. 
By Eq.~\eqref{vn2DecayCorr}, its magnitude is determined by two competing factors: it is suppressed by a factor $v_n^2$, meanwhile, augmented by a factor of $N$.

The above calculations can be extended to $\langle\langle 4\rangle\rangle$ for the correlated case, where more cross terms contribute to the resulting expression.
After somewhat tedious calculations, one finds
\bqn
\langle\langle 4 \rangle\rangle &=& v_n^4\frac{P(N, 4)}{P(N+2M, 4)} \nb\\
&&+ \frac12\left(1+\cos (n\phi_\mathrm{open})\right)v_n^4\frac{4 P(N, 3)P(2M, 1)}{P(N+2M, 4)} \nb\\
&&+ \left[\frac23\cos (n\phi_\mathrm{open})v_n^2 
+ \frac13\times 0\right] \frac{6 P(N,2)2M}{P(N+2M, 4)} \nb\\
&&+ \left[\frac23\frac12\left(1+\cos (n\phi_\mathrm{open})\right)v_n^4 
+ \frac13 \frac14\left(1 + 2\cos (n\phi_\mathrm{open}) + (2\cos^2 (n\phi_\mathrm{open})-1)\right)v_n^4\right] \frac{6 P(N,2)4P(M,2)}{P(N+2M, 4)} \nb\\
&&+ \frac{1}{4\times 2^3}\left[4 + 12\cos (n\phi_\mathrm{open}) +12\left(\frac23\times 1+\frac13(2\cos^2 (n\phi_\mathrm{open})-1)\right) +4\cos (n\phi_\mathrm{open})\right] v_n^4\frac{4 P(N, 1)8P(M,3)}{P(N+2M, 4)} \nb\\
&&+ \left[\frac23\frac14\left((2\cos^2 (n\phi_\mathrm{open})-1)+2\cos (n\phi_\mathrm{open})+1\right)v_n^2
+\frac13 \frac12\left(0+0\right)\right]\frac{4 P(N, 1)12P(M,2)}{P(N+2M, 4)} \nb\\
&&+ \left[\frac23\frac12\left(2\cos^2 (n\phi_\mathrm{open})-1+1\right)+\frac13\times 0\right]\frac{12P(M, 2)}{P(N+2M, 4)} \nb\\
&&+ \left[\frac23\frac14\left((2\cos^2 (n\phi_\mathrm{open})-1)+2\cos(n\phi_\mathrm{open})+1\right)v_n^2
+\frac13 \times 0\right]\frac{48P(M, 3)}{P(N+2M, 4)} \nb\\
&&+ \frac{1}{2^4 4!}\left(144 +192\cos (n\phi_\mathrm{open}) + 48\left(2\cos^2 (n\phi_\mathrm{open})-1\right)\right) v_n^4 \frac{16P(M, 4)}{P(N+2M, 4)} \nb\\
&\equiv& C_{0} + C_{1} \cos(n\phi_\mathrm{open}) + C_{2} \cos^2(n\phi_\mathrm{open})  ,
\label{bra4Corr}
\eqn
where
\begin{eqnarray}
C_{0} &=& \frac{v_n^4}{P(N+2M,4)}\left[P(N,4)+4M P(N,3)+8 P(N,2)P(M,2)+8 P(N,1)P(M,3)+4 P(M,4)\right] , \nonumber\\
C_{1} &=& \frac{v_n^2}{P(N+2M,4)}\left[8M P(N,2)+16P(N,1)P(M,2)+16 P(M,3)\right]\nb\\
&&+\frac{v_n^4}{P(N+2M,4)}\left[4M P(N,3) +12 P(N,2)P(M,2)+16 P(N,1)P(M,3)+8 P(M,4)\right], \nonumber\\
C_{2} &=& \frac{1}{P(N+2M,4)}8P(M,2)\nb\\
&&+\frac{v_n^2}{P(N+2M,4)}\left[16P(N,1)P(M,2)+16 P(M,3)\right]\nb\\
&&+\frac{v_n^4}{P(N+2M,4)}\left[4 P(N,2)P(M,2)+8 P(N,1)P(M,3)+4 P(M,4)\right]  .
\end{eqnarray}
In the derivation, besides Eq.~\eqref{PN+2M4}, one invokes the following combination numbers $P(2M,4)=3\times 2^2P(M,2)+3\times 2\times 2\times 2^2P(M,3)+2^4P(M,4)$, $P(2M,3)=2^3 P(M,3)+3\times 2^2 P(M,2)$, $P(2M,2)=2^2 P(M,2)+2 P(M,1)$, $4\times 2^3=4+12+12+4$, and $2^4 4!=24+96+144+96+24$ that are related to different contributions.
The complexity primarily lies in the fact that, due to the correlation with the symmetry plane, every single combination yields a non-vanishing but mostly distinct contribution.
As an exmple, consider the contribution 
\bqn
\frac{1}{4\times 2^3}\times 12\left[\frac23\times 1+\frac13\times(2\cos^2 (n\phi_\mathrm{open})-1)\right]  v_n^4\frac{4 P(N, 1)8P(M,3)}{P(N+2M, 4)} \nb
\eqn
on the 5th line of Eq.~\eqref{bra4Corr}.
The factor $4 P(N, 1)8P(M,3)$ is the combinatorial number of picking out one particle from the background and three from three different pairs.
Considering that the particle from the pairs might be either the first or the second particle forming the pair, the total permutation number of the four particles is $4\times 2^3$, where $4$ stand for possible positions of the background particle in the four-particle tuple.
Out of $4\times 2^3$ possibilities, there are $12$ cases that two particles are the second particle of the pair in question, among which two third chance the azimuthal angle of these two particle have the opposite signs in the correlator $\langle\langle4\rangle\rangle$ and one third they have the same sign.
In the former case, one involves the integral
\bqn
&&\frac{1}{(2\pi)^4}\int d\phi_1 d\phi_2 d\phi_3 d\phi_4 \cos n(\phi_1-\phi_2+\phi_3-\phi_4) \nb\\
&&~~~~(1+2v_n\cos n(\phi_1+\phi_\mathrm{open}-\Psi_n)) (1+2v_n\cos n(\phi_2+\phi_\mathrm{open}-\Psi_n)) (1+2v_n\cos n(\phi_3-\Psi_n)) (1+2v_n\cos n(\phi_4-\Psi_n)) \nb\\
&&= v_n^4 ,\nb
\eqn
and for the latter, one makes use of the result
\bqn
&&\frac{1}{(2\pi)^4}\int d\phi_1 d\phi_2 d\phi_3 d\phi_4 \cos n(\phi_1+\phi_2-\phi_3-\phi_4) \nb\\
&&~~~~(1+2v_n\cos n(\phi_1+\phi_\mathrm{open}-\Psi_n)) (1+2v_n\cos n(\phi_2+\phi_\mathrm{open}-\Psi_n)) (1+2v_n\cos n(\phi_3-\Psi_n)) (1+2v_n\cos n(\phi_4-\Psi_n)) \nb\\
&&= v_n^4 (2\cos^2 (n\phi_\mathrm{open})-1) .\nb
\eqn
The remainder of Eq.~\eqref{bra4Corr} can be derived using similar rationale.

Using Eq.~\eqref{cn2DecayCorr} and carrying out the same expansion for $2\langle\langle 2\rangle\rangle^2$, one finds
\bqn
\widetilde{C}_{0} &=& \frac{v_n^4 }{P(N+2M,2)^2}2\left[P(N+M,2)+P(M,2)\right]^2, \nb\\
\widetilde{C}_{1} &=& \frac{ v_n^2 }{P(N+2M,2)^2}8M\left[P(N+M,2)+P(M,2)\right]
+\frac{v_n^4 }{P(N+2M,2)^2}8M(N+M-1)\left[P(N+M,2)+P(M,2)\right], \nb\\
\widetilde{C}_{2} &=& \frac{1}{P(N+2M,2)^2}8M^2
+\frac{v_n^2}{P(N+2M,2)^2}16 M^2 (N+M-1)
+\frac{v_n^4}{P(N+2M,2)^2}8M^2(N+M-1)^2 .
\eqn
As one evaluates $2\langle\langle 2\rangle\rangle^2 - \langle\langle 4\rangle\rangle$, one finds a strong cancellation in most, but not all, of the leading terms.  
As mentioned previously, this cancellation can be largely attributed to the definition of the four-particle correlator, which is designed to eliminate contributions arising from two-particle correlations.  
However, because pair emission is also correlated with the symmetry plane, this cancellation is not entirely perfect from an analytical standpoint.  
Specifically, for $C_0$ and $\widetilde{C}_0$, the leading contributions are  
\bqn
\widetilde{C}_0 \sim 2C_0 \sim \frac{v_n^{4}}{(N+2M)^{4}}\left(N^{4}+4MN^{3}+8N^{2}M^{2}+8NM^{3}+4M^{4}\right) .
\eqn
For $C_1$ and $\widetilde{C}_1$, the leading contributions proportional to $v_n^2$ are identical, namely,  
\bqn
\widetilde{C}_1 \sim C_1 \sim \frac{v_n^{2}}{(N+2M)^{4}}\left(8MN^{2}+16NM^{2}+16M^{3}\right) .
\eqn
For $C_2$ and $\widetilde{C}_2$, the leading contributions from the constant and those proportional to $v_n^2$ are identical, namely,  
\bqn
\widetilde{C}_2 \sim C_2 \sim \frac{1}{(N+2M)^{4}}\left[8M^{2}+v_n^{2}\left(16NM^{2}+16M^{3}\right)\right] .
\eqn
As a result, the residual in $2\langle\langle 2\rangle\rangle^2 - \langle\langle 4\rangle\rangle$ is primarily governed by a form proportional to $v_n^4$, with subleading terms supressed by a factor of $1/v_n^2(N+M)$. 

Putting all the pieces together we find $v_n\{4\}$ up to the subleading contributions
\bqn
v_n\{4\} &=& \left\{\left[v_n^2\frac{\left(P(N+M,2)+P(M,2)\right)+ 2\cos\left(n\phi_{\mathrm{open}}\right)M(N+M-1) }{P(N+2M,2)}\right]^2\right.\nb\\
&&\left.-v_n^2\left[\cos\left(n\phi_{\mathrm{open}}\right)\frac{32 \left(MN^2+2NM^2+2M^3\right)}{(N+2M)^5}
+\cos^2\left(n\phi_{\mathrm{open}}\right)\frac{64 M^2 (N+M)}{(N+2M)^5}\right]
\right\}^\frac14 .
\label{vn4DecayCorr}
\eqn
The leading term on the first line of Eq.~\eqref{vn4DecayCorr} is largely identical to Eq.~\eqref{vn2DecayCorr} up to a shift $\frac{2M}{P(N+2M,2)}\cos\left(n\phi_{\mathrm{open}}\right)$, which is readily attributed to the strong cancellation discussed above.
It is worth noting that the persistence of dependence on $\cos\left(n\phi_{\mathrm{open}}\right)$, a measure of two-particle correlation of the emitted pairs, is a manifestation that the resulting four-particle correlation still hings on the particle pair due to their correlation with the symmetry plane. 
It is noted that the performance of $v_n\{6\}$ is found numerically similar to that of $v_n\{4\}$, this is understood as there are no intrinsic three-particle correlations in the toy model.  

\begin{figure}[ht]
    \centering
    \begin{minipage}{0.4\textwidth}
        \centering
        \includegraphics[width=1.0\textwidth, height=0.23\textheight]{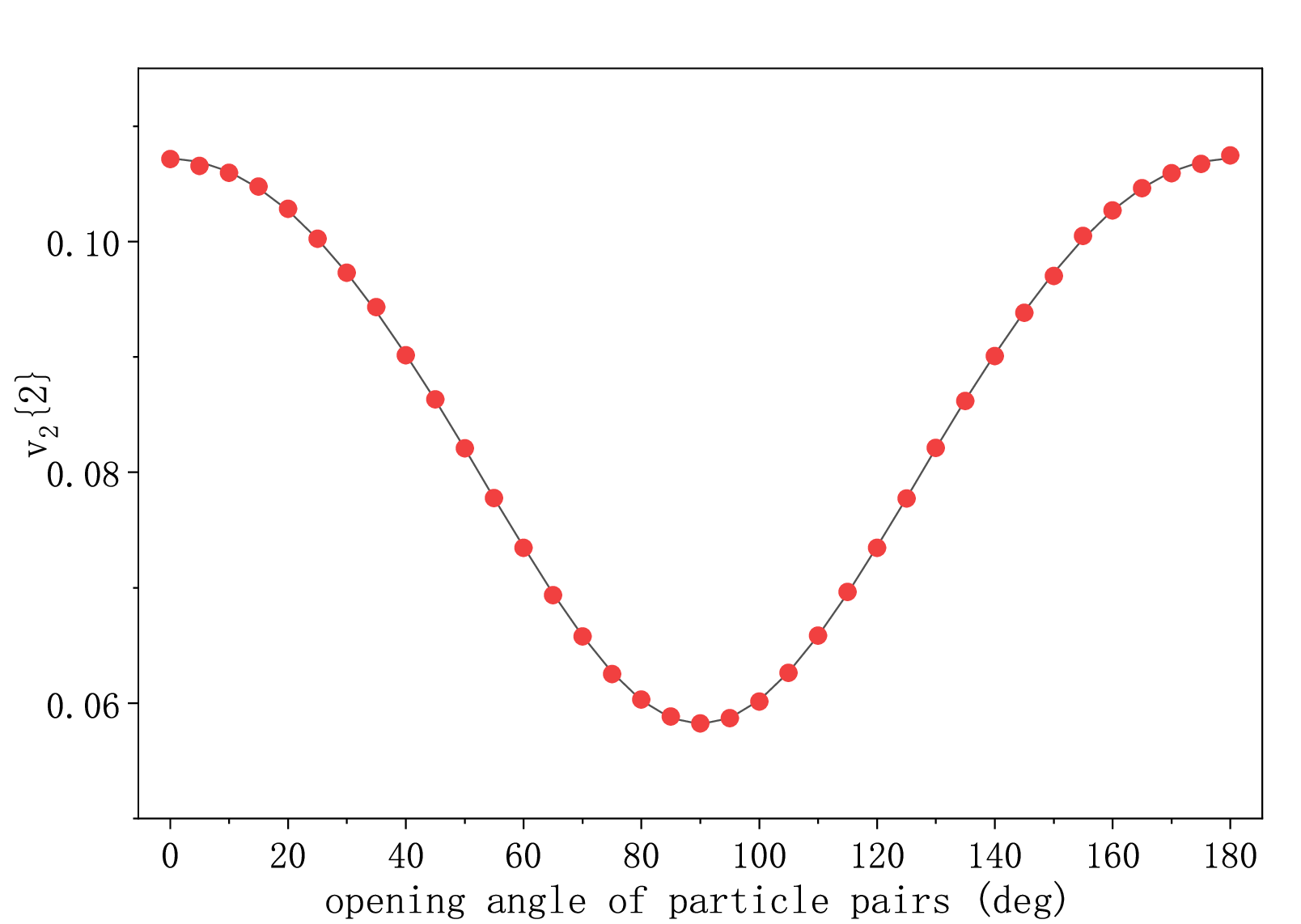}
    \end{minipage}
    \begin{minipage}{0.4\textwidth}
        \centering
        \includegraphics[width=1.0\textwidth, height=0.23\textheight]{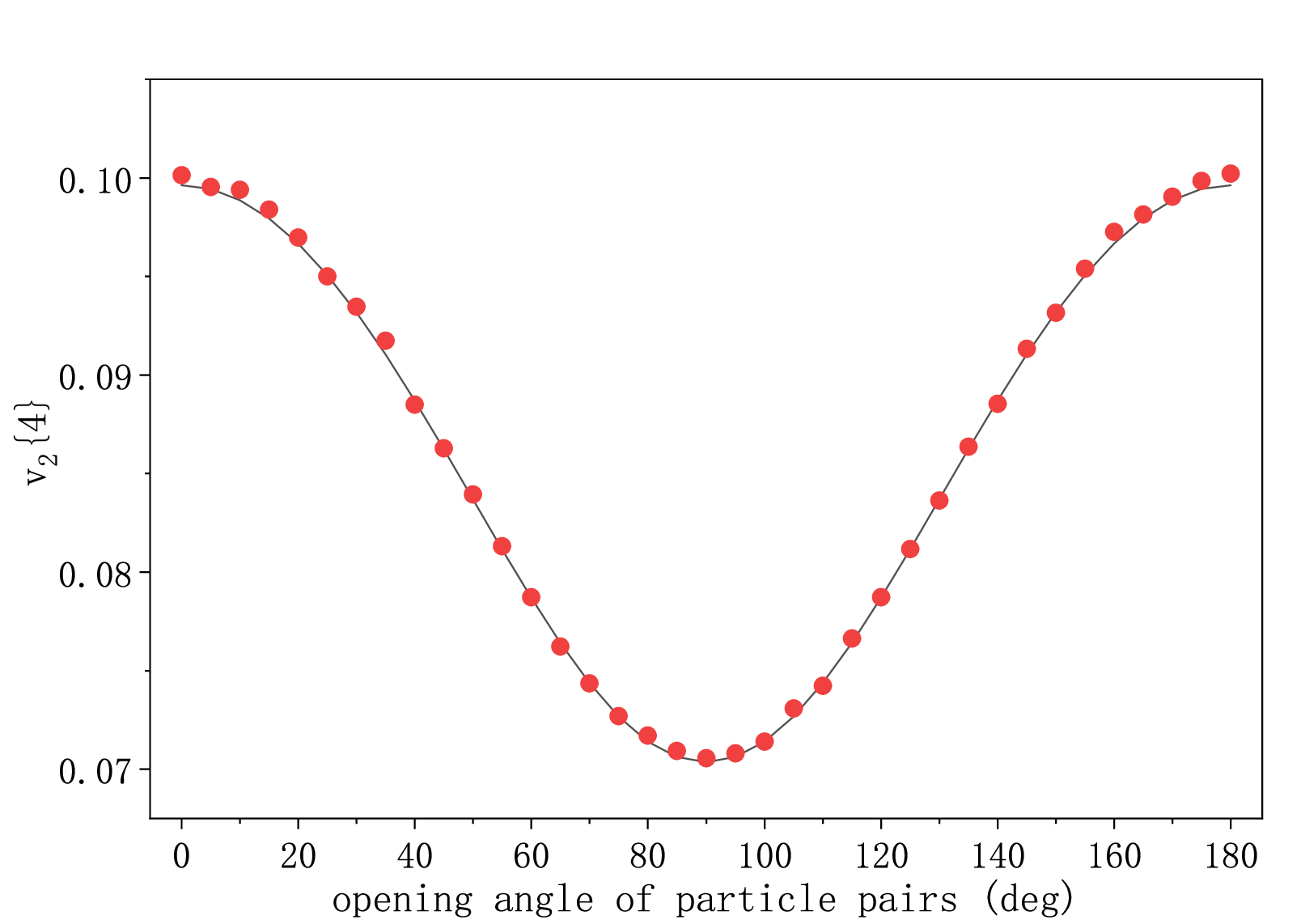}
    \end{minipage}
    \begin{minipage}{0.4\textwidth}
        \centering
        \includegraphics[width=1.0\textwidth, height=0.23\textheight]{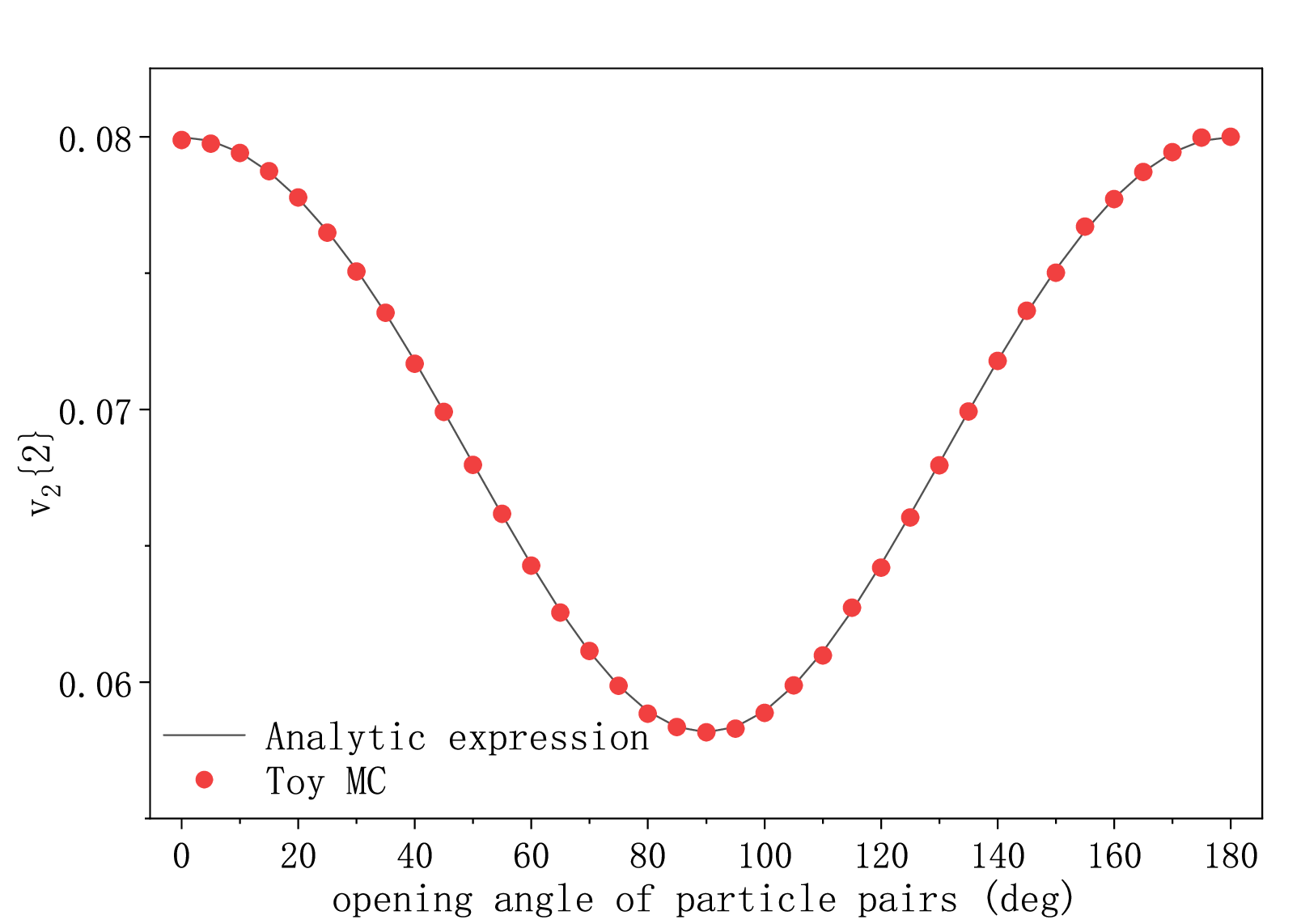}
    \end{minipage}
    \begin{minipage}{0.4\textwidth}
        \centering
        \includegraphics[width=1.0\textwidth, height=0.23\textheight]{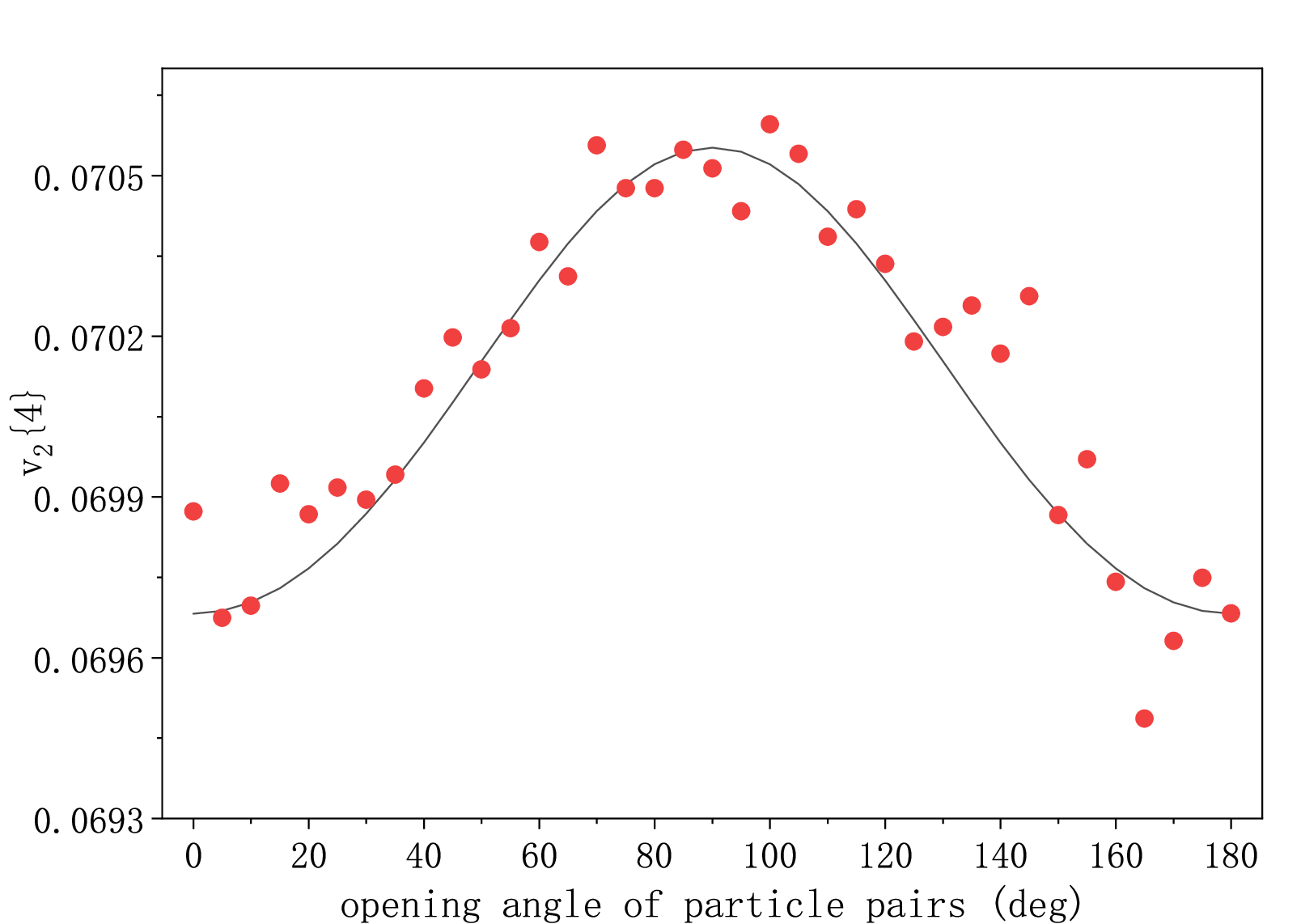}
    \end{minipage}
\renewcommand{\figurename}{Fig.}   
\caption{Comparison between the analytic expressions and the numerical results for the elliptic flow harmonics $v_2\{2\}$ and $v_2\{4\}$ as functions of the opening angle $\phi_{\mathrm{open}}$ in toy model I.
The upper row corresponds to events in which the emitted pairs are correlated with the symmetry plane, whereas the lower row represents the uncorrelated case.
The analytic curves are obtained using Eqs.~\eqref{vn2DecayCorr},~\eqref{vn4DecayCorr},~\eqref{vn2DecayUncorr}, and~\eqref{vn4DecayUncorr}, respectively.
The numerical simulations are carried out by assuming the input flow harmonics $v_2=0.1$ and $v_3=0.06$ for a total of 50,000 events with $M = 30$ emitted particle pairs and a total multiplicity $N + 2M = 200$.}
\label{v2v4_analysis}
\end{figure}

These analytic results are in good agreement with those obtained from the numerical simulations presented in the previous section.
In Fig.~\ref{v2v4_analysis}, we confront the derived expressions with the numerical results for $v_2\{2\}$ and $v_2\{4\}$ as functions of the opening angle $\phi_{\mathrm{open}}$.
The upper row displays events in which the emitted pairs are aligned with the symmetry plane, whereas the lower row corresponds to events where the pairs are uncorrelated with it.
The analytic curves are obtained from Eqs.~\eqref{vn2DecayCorr},~\eqref{vn4DecayCorr},~\eqref{vn2DecayUncorr}, and~\eqref{vn4DecayUncorr}, respectively.
The numerical data are generated from 50,000 Monte Carlo events with $M = 30$ emitted particle pairs and a total multiplicity of $N + 2M = 200$.
Overall, the simulations reproduce the analytic predictions quite well in all four panels.
For the case with pair emission correlated with the symmetry plane, shown in the upper row, both the magnitude and the dependence on the opening angle are accurately described by the analytic formulas.
In the lower row, representing uncorrelated pair emission, the same level of agreement is observed.
In the bottom-right panel, the description of $v_2\{4\}$ remains satisfactory, although the numerical points exhibit noticeable scatter due to larger statistical fluctuations.
This behavior can be understood by substituting the specific values $N = 140$, $M = 30$, and $v_2=0.1$ into the subleading oscillatory terms of Eqs.~\eqref{vn4DecayCorr} and~\eqref{vn4DecayUncorr}.
One finds that the coefficients in front of $\cos(n\phi_{\mathrm{open}})$ take the values $3.70\times 10^{-5}$ and $-5.88\times 10^{-7}$, respectively, in the correlated and uncorrelated cases.
The smallness of the latter accounts for the more pronounced scatter observed in the numerical simulations in the bottom-right panel.

All in all, while converging to the apparent value, particle correlators underestimate the input flow harmonics, if the latter is the underlying information to extracted.
Such an underestimation of the particle correlator with respect to input flow can partly be attributed to the ``collective'' nature of particle decay, which typically affects a given fraction of the total particles, and therefore the suppression of the high-order correlator may not be effective as expected.
Nonetheless, this is not the case for toy model II, indicating that the particular feature of the correlators in the context of global momentum conservation is an interesting topic to explore.

\subsection{Toy model II}

For toy model II, as one assumes that all the emitted particles have the same modulus $p_\mathrm{T}=1$, we have the following form for the distribution function of an event consisting of $N$-particles:
\bqn
f_N(\phi_1,\ldots,\phi_N) =
\frac{f(\phi_1),\ldots,f(\phi_j)\delta\left(\sum_{j=1}^N e^{i\phi_j}\right)}
{\mathlarger{\int} d\phi_1\ldots d\phi_N\, \delta\left(\sum_{j=1}^N e^{i\phi_j}\right)} ,\label{NParDis}
\eqn
where the factor of $\delta$-function enforces the momentum conservation and $f\equiv f_1$ is the original one-particle distribution function Eq.~\eqref{oneParDis}.
The phase space integration on the denominator is essential a normalization factor owing to the global momentum conservation.
We note that it does not change the background flow defined by Eq.~\eqref{defvn}.

To see this, we note that $k$-particle distribution of the above $N$-particle event reads
\bqn
f_k(\phi_1,\ldots,\phi_k) =
\frac{f(\phi_1),\ldots,f(\phi_k)\mathlarger{\int} d\phi_{k+1}\ldots d\phi_N\, f(\phi_{k+1}),\ldots,f(\phi_N)\,\delta\left(\sum_{j=1}^N e^{i\phi_j}\right)}
{\mathlarger{\int} d\phi_1\ldots d\phi_N\, \delta\left(\sum_{j=1}^N e^{i\phi_j}\right)} ,\label{kParDis}
\eqn
which, in practice, can be approximated either by invoking the central limit theorem when if $N-k$ is large~\cite{hydro-corr-non-flow-01} or rewriting the $\delta$-function as an exponential integral and evaluating it using the saddle-point approximation~\cite{hydro-corr-non-flow-02}. 
For the purposes of the present study, Eq.~\eqref{kParDis} implies that the ``modified'' one-particle distribution function takes the form
\bqn
f_1(\phi_1) =
\frac{f(\phi_1)\mathlarger{\int} d\phi_{2}\ldots d\phi_N\, f(\phi_2),\ldots,f(\phi_N)\,\delta\left(\sum_{j=1}^N e^{i\phi_j}\right)}
{\mathlarger{\int} d\phi_1\ldots d\phi_N\, \delta\left(\sum_{j=1}^N e^{i\phi_j}\right)}
\equiv f(\phi_1)\mathcal{N}_1(\phi_1) .
\eqn
Now we argue that the correction $\mathcal{N}_1(\phi)$
\bqn
\mathcal{N}_1(\phi_1) \equiv \frac{\mathlarger{\int} d\phi_{2}\ldots d\phi_N\, f(\phi_2),\ldots,f(\phi_N)\,\delta\left(\sum_{j=1}^N e^{i\phi_j}\right)}
{\mathlarger{\int} d\phi_1\ldots d\phi_N\, \delta\left(\sum_{j=1}^N e^{i\phi_j}\right)} \label{DefCalN1}
\eqn
is isotropic.
This is because $\mathcal{N}_1(\phi+\Delta \phi) =\mathcal{N}_1(\phi)$, which is readily seen by a translation of the variables $\phi_j\to\phi_j+\Delta\phi$ in Eq.~\eqref{DefCalN1} while noticing $\delta\left(\sum_{j=1}^N e^{i\phi_j}\right)=\delta\left(\sum_{j=1}^N e^{i(\phi_j+\Delta\phi)}\right)$.
Furthermore, the normalization of $f_1$ dictates that it must be a trivial constant, namely, $\mathcal{N}_1(\phi)=1$, while non-trivial result only appears in $\mathcal{N}_k$ for $k\ge 2$.
From the above result, one finds that the flow harmonics 
\bqn
v_n = \mathlarger{\int} d\phi_1 \cos n(\phi_1-\Psi_n) f(\phi_1) \mathcal{N}_1(\phi_1)
\eqn
remains intact under global momentum conservation.

We argue that the above seemingly surprising result admits an intuitive justification.
First, consider the simplest possible scenario where the original one-particle distribution $f_1(\phi)$ is isotropic.
Since momentum conservation favors no particular azimuthal direction, the modified one-particle distribution must remain isotropic.
Now consider a general one-particle distribution with non-vanishing flow harmonics.
The situation remains unchanged: momentum conservation is azimuthally independent and possesses no non-trivial azimuthal symmetry.
In other words, it neither prefers any azimuthal direction nor correlates with any symmetry plane.
Again, there is no reason for it to alter the azimuthal dependence of the modified one-particle distribution.
However, as the conservation law manifests as an interaction between particles, the quantities that are indeed affected are the multi-particle azimuthal correlators, as evident in the modifications to $v_1\{2\}$~\cite{hydro-corr-non-flow-01, hydro-corr-non-flow-02}, $v_2\{2\}$, and $v_2\{4\}$~\cite{hydro-corr-non-flow-14, hydro-corr-non-flow-20}.
In this regard, as flow estimators, multi-particle correlators themselves introduce an artificial effect that can lead to their underperformance, an effect absent in single-particle flow estimators such as the event-plane and MLE methods.
Furthermore, one can provide an explanation for the relation in Eq.~\eqref{cnAssEV}, obtained from the event-plane method, which is distinct from the one extracted from particle correlators.
Specifically, for an isotropic distribution, even though expected value vanishes $\langle \cos n\phi \rangle = 0$, the average over $M$ emitted particles,
\begin{equation}
\overline{\cos n\phi}
= \frac{1}{M}\sum_{i=1}^{M} \cos(n\phi_i) \, ,
\end{equation}
follows a normal distribution with standard deviation $\sigma/\sqrt{M}$, where the variance is given by $\sigma^2 = \langle \cos^2 n\phi \rangle = 1/2$, as a consequence of the central limit theorem.
From this observation, the relation in Eq.~\eqref{cnAssEV}, as displayed in Fig.~\ref{cmp_scheme2}, is readily obtained.

\section{Concluding remarks}\label{section5}

Over the past decade, flow measurements in high-multiplicity proton-proton, proton-nucleus, and other small collision systems at RHIC and the LHC have led to a renewed interest in non-flow suppression. 
In such systems, short-range correlations from jets, resonance decays, and other few-body processes can be comparable in magnitude to the genuine collective signal, posing a challenge for flow extraction and interpretation. 
This has motivated the development and systematic application of analysis strategies such as multi-particle cumulants, sub-event cumulant methods, and large pseudorapidity gaps, which are explicitly designed to reduce the sensitivity of flow observables to non-flow contributions in small systems. 
Amid these efforts, multi-particle correlators, especially higher-order ones, are generally regarded as robust flow estimators whose construction strongly suppresses non-flow effects by combinatorial arguments. 
The standard picture is that non-flow arises primarily from few-particle correlations, so its relative weight decreases rapidly as the order of the correlator increases, while the collective flow signal, being a genuine many-body phenomenon, continues to contribute coherently. 
This expectation underlies the common practice of using multi-particle correlators as benchmarks for collectivity in small systems and of interpreting sign changes or magnitude differences between, for example, $v_2\{2\}$ and $v_2\{4\}$ as indicators of the interplay between flow and non-flow. 

The present study revisits this general understanding using two controlled toy models that emulate typical sources of non-flow, namely particle decay and global momentum conservation, and probes the performance of multi-particle correlators as flow estimators in these settings. 
In the first toy model that mimics non-flow due to the particle decay process, we find that higher-order particle correlators, such as $v_2\{4\}$ and $v_2\{6\}$, generally follow the apparent flow harmonics of the final distorted particle distribution more closely than lower-order estimators, while they do not necessarily provide a better reconstruction of the input background flow before the non-flow modification. 
Their behavior depends sensitively on the opening angle of the emitted particle pairs and on whether the pair emission is correlated with the symmetry plane, leading to characteristic oscillations and suppressions that can be understood analytically.
Moreover, we provide an analytic account of the characteristic oscillations observed in $v_2\{2\}$ and $v_2\{4\}$, showing that the response of these estimators to the injected non-flow correlations can be more intricate than suggested by simple suppression arguments. 
In the second toy model simulating the impact of global momentum conservation, we analyze how the induced long-range constraints modify the behavior of multi-particle observables. 
Analytically, we show that multi-particle correlators can generate an artificial contribution that is absent in other estimators, revealing a specific sensitivity to this particular class of non-flow that is unique to the cumulant-based framework.

Taken together, these results indicate that, especially in small or low-multiplicity systems where non-flow is comparatively large, the apparent robustness of multi-particle correlators depends sensitively on the structure of the underlying non-flow mechanism and on the quantity one aims to reconstruct. 
In particular, the present study focuses on situations in which higher-order cumulants may remain close to the apparent flow harmonics of the final distorted particle distribution, while not necessarily providing a better reconstruction of the input background flow. 
The present study therefore suggests that suppressing explicit few-particle non-flow in a correlator does not always automatically imply a faithful reconstruction of the underlying input flow harmonics. 
We plan to continue investigating this topic in future studies.

\section*{Acknowledgements}

We are thankful for the enlightening discussions with Mike Lisa, Hong-Hao Ma, Sandra Padula, and Cesar Bernardes for insightful discussions.
The authors are deeply indebted to Yogiro Hama for his inspiring guidance and unwavering encouragement throughout the years.
We gratefully acknowledge the financial support from Brazilian agencies 
Funda\c{c}\~ao de Amparo \`a Pesquisa do Estado de S\~ao Paulo (FAPESP), 
Funda\c{c}\~ao de Amparo \`a Pesquisa do Estado do Rio de Janeiro (FAPERJ), 
Conselho Nacional de Desenvolvimento Cient\'{\i}fico e Tecnol\'ogico (CNPq), 
and Coordena\c{c}\~ao de Aperfei\c{c}oamento de Pessoal de N\'ivel Superior (CAPES).
A part of this work was developed under the project Institutos Nacionais de Ciências e Tecnologia - Física Nuclear e Aplicações (INCT/FNA) Proc. No. 408419/2024-5.
This research is also supported by the Center for Scientific Computing (NCC/GridUNESP) of São Paulo State University (UNESP).
CY acknowledges the support of the FAPERJ process no. E-26/200.231/2025.

\bibliographystyle{h-physrev}
\bibliography{references_qian}

\end{document}